\documentclass{article}
\usepackage{arxiv}

\usepackage[utf8]{inputenc}
\usepackage[T1]{fontenc}
\usepackage[hidelinks]{hyperref}
\usepackage{url}
\usepackage{booktabs}
\usepackage{tabularx}
\usepackage{amsmath,amsfonts,amssymb}
\usepackage{microtype}
\usepackage{graphicx}
\usepackage{afterpage}
\usepackage[super,numbers,sort&compress]{natbib}   
\usepackage{xcolor}

\newcommand{\nDaicPersons}{189}
\newcommand{\nDaicWindows}{10,533}
\newcommand{\nDaicQuestions}{108}
\newcommand{\nDaicWinPerPerson}{55}
\newcommand{\nDaicWordsPerWin}{12}
\newcommand{\nEdaicPersons}{86}
\newcommand{\nEdaicWindows}{5,214}
\newcommand{\nEdaicWinPerPerson}{60}
\newcommand{\nEdaicWordsPerWin}{8}
\newcommand{\nAllPersons}{275}
\newcommand{\nAlexSessions}{786}
\newcommand{\nAlexTurns}{180,547}

\newcommand{\nAnnomiSessions}{133}

\newcommand{\nHlqcSessions}{258}

\newcommand{\nHlqcShared}{74}
\newcommand{\nHlqcIndep}{184}
\newcommand{\nAnnomiHigh}{110}
\newcommand{\nAnnomiLow}{23}

\newcommand{\nHlqcIndepHigh}{119}
\newcommand{\nHlqcIndepLow}{65}
\newcommand{\nSymItems}{25}

\newcommand{\nWaiReverse}{14}
\newcommand{\nNliAtoms}{42}
\newcommand{\nNliNeutral}{\ensuremath{-}0.19}
\newcommand{\nNliKdiag}{0.45}
\newcommand{\nNliKoff}{\ensuremath{-}0.07}
\newcommand{\nWordRankMini}{6.0}
\newcommand{\nWordRankBge}{2.6}
\newcommand{\nDupSleep}{0.90}
\newcommand{\nDupConc}{0.73}
\newcommand{\nDupInterest}{0.64}
\newcommand{\nWAICFF}{0.55}
\newcommand{\nWAICFR}{0.55}

\newcommand{\nSimTruthWording}{0.50}
\newcommand{\nSimNaiveZero}{0.47}

\newcommand{\nSimWhiteZero}{0.01}
\newcommand{\nSimWhiteFour}{0.44}

\newcommand{\nSimWhiteBeyondFour}{0.21}
\newcommand{\nSimNaiveBeyondFour}{0.08}

\newcommand{\nDMiniSrWord}{0.59}
\newcommand{\nDMiniNaiveSr}{0.42}
\newcommand{\nDMiniNaiveSrPartial}{0.03}

\newcommand{\nDBgeNaiveSr}{0.47}
\newcommand{\nDBgeNaiveSrPartial}{0.12}
\newcommand{\nCMiniSrWord}{0.43}
\newcommand{\nCMiniNaiveSr}{0.36}
\newcommand{\nCMiniNaiveSrPartial}{0.13}

\newcommand{\nCBgeNaiveSr}{0.21}
\newcommand{\nCBgeNaiveSrPartial}{\ensuremath{-}0.07}
\newcommand{\nHubEmbDE}{0.92}
\newcommand{\nHubEmbWordD}{0.74}
\newcommand{\nHubEmbWordE}{0.72}
\newcommand{\nHubEmbSelfD}{0.43}
\newcommand{\nHubEmbSelfE}{0.32}
\newcommand{\nHubSelfDE}{0.56}
\newcommand{\nHubSelfWordD}{0.50}
\newcommand{\nHubSelfWordE}{0.35}
\newcommand{\nHubEntWordD}{0.29}

\newcommand{\nHubEntDE}{0.78}
\newcommand{\nHubEntProbeD}{0.34}
\newcommand{\nHubEntProbeE}{0.36}
\newcommand{\nHubTheoryMax}{0.31}
\newcommand{\nMethodRScore}{0.54}
\newcommand{\nRelEmbD}{0.95}

\newcommand{\nRelSelfD}{0.68}
\newcommand{\nRelSelfE}{0.46}
\newcommand{\nHubEmbSelfDNorm}{0.53}
\newcommand{\nHubEmbWordDNorm}{0.75}

\newcommand{\nHubTauAgree}{0.98}
\newcommand{\nMethodRWording}{0.61}
\newcommand{\nStanceCosFwd}{0.68}
\newcommand{\nStanceCosRev}{0.71}
\newcommand{\nStanceNliFwd}{0.09}
\newcommand{\nStanceNliRev}{\ensuremath{-}0.99}
\newcommand{\nFLNullN}{7.5}
\newcommand{\nFLNullS}{5.0}
\newcommand{\nFLPowerFourN}{94}
\newcommand{\nFLPowerTwoN}{15}

\newcommand{\nFLReps}{200}

\newcommand{\nSimCosTask}{0.01}

\newcommand{\nSimTalkTask}{0.96}
\newcommand{\nSimCosBond}{0.05}

\newcommand{\nSimTalkBond}{0.98}

\newcommand{\nSimStanceGoal}{0.82}

\newcommand{\nDaicEtaQ}{0.110}
\newcommand{\nDaicEtaQlo}{0.058}
\newcommand{\nDaicEtaQhi}{0.150}
\newcommand{\nDaicEtaP}{0.039}
\newcommand{\nDaicRWindow}{0.89}
\newcommand{\nDaicRPerson}{0.68}
\newcommand{\nDaicRPersonCentred}{0.63}
\newcommand{\nDaicRWhitened}{0.40}
\newcommand{\nDaicRelWhite}{0.88}

\newcommand{\nDaicRelDev}{0.05}

\newcommand{\nEdaicEtaQ}{0.071}

\newcommand{\nEdaicEtaP}{0.030}

\newcommand{\nEdaicRelWhite}{0.93}

\newcommand{\nEdaicRelDev}{0.03}

\newcommand{\nDMiniHoneW}{0.68}
\newcommand{\nDMiniHoneS}{0.42}
\newcommand{\nDMiniHoneD}{0.26}
\newcommand{\nDMiniHoneCI}{0.15 to 0.47}
\newcommand{\nDMiniHtwoR}{\ensuremath{-}0.009}
\newcommand{\nDMiniHtwoP}{0.93}

\newcommand{\nDBgeHoneW}{0.69}
\newcommand{\nDBgeHoneS}{0.47}
\newcommand{\nDBgeHoneD}{0.22}
\newcommand{\nDBgeHoneCI}{0.15 to 0.40}
\newcommand{\nDBgeHtwoR}{0.028}
\newcommand{\nDBgeHtwoP}{0.52}

\newcommand{\nDNliHtwoR}{0.051}
\newcommand{\nDNliHtwoP}{0.27}

\newcommand{\nDNliHthreeD}{0.038}
\newcommand{\nDNliHthreeCI}{0.023 to 0.056}
\newcommand{\nCMiniHoneW}{0.63}
\newcommand{\nCMiniHoneS}{0.36}
\newcommand{\nCMiniHoneD}{0.27}
\newcommand{\nCMiniHoneCI}{0.11 to 0.55}
\newcommand{\nCMiniHtwoR}{0.008}

\newcommand{\nCMiniHtwoPholm}{0.93}

\newcommand{\nCMiniHthreeD}{0.038}
\newcommand{\nCMiniHthreeCI}{0.007 to 0.072}
\newcommand{\nCBgeHoneW}{0.65}
\newcommand{\nCBgeHoneS}{0.21}
\newcommand{\nCBgeHoneD}{0.44}
\newcommand{\nCBgeHoneCI}{0.33 to 0.58}
\newcommand{\nCBgeHtwoR}{\ensuremath{-}0.096}

\newcommand{\nCBgeHtwoPholm}{0.15}

\newcommand{\nCBgeHthreeD}{0.042}
\newcommand{\nCBgeHthreeCI}{0.013 to 0.073}
\newcommand{\nCNliHtwoR}{0.118}

\newcommand{\nCNliHtwoPholm}{0.15}

\newcommand{\nCNliHthreeD}{0.005}
\newcommand{\nCNliHthreeCI}{\ensuremath{-}0.022 to 0.035}
\newcommand{\nDMiniG}{0.07}

\newcommand{\nDMiniSpec}{0.005}

\newcommand{\nDBgeG}{0.26}

\newcommand{\nDNliG}{0.49}
\newcommand{\nDNliGCI}{0.38 to 0.59}

\newcommand{\nDNliMeanPHQ}{0.53}
\newcommand{\nCMiniG}{0.33}

\newcommand{\nCBgeG}{0.52}

\newcommand{\nCBgeSpec}{0.073}

\newcommand{\nCNliG}{0.60}
\newcommand{\nCNliGCI}{0.44 to 0.72}

\newcommand{\nCNliMeanPHQ}{0.58}
\newcommand{\nCompSessions}{786}
\newcommand{\nCompBoth}{783}
\newcommand{\nCompMulti}{159}
\newcommand{\nCompAnx}{498}
\newcommand{\nCompDep}{374}
\newcommand{\nCompScz}{71}
\newcommand{\nCompSui}{12}
\newcommand{\nCompClTaskRsq}{0.06}

\newcommand{\nCompClBondRsq}{0.91}

\newcommand{\nCompThBondRsq}{0.66}

\newcommand{\nCompEtaCollMax}{0.02}
\newcommand{\nCompGapBond}{0.94}

\newcommand{\nRdocDefs}{49}
\newcommand{\nRdocDomains}{6}
\newcommand{\nRdocNNMini}{0.82}

\newcommand{\nRdocChance}{0.18}

\newcommand{\nMbItems}{702}
\newcommand{\nMbBanks}{35}

\newcommand{\nMbBgeRawMin}{1.3}
\newcommand{\nMbBgeRawMax}{2.9}
\newcommand{\nMbBgeMRMax}{16.8}

\newcommand{\nMbNNDomMedian}{0.05}

\newcommand{\nMbInstruments}{34}
\newcommand{\nMbNNDomTIPI}{0.70}
\newcommand{\nMbNNDomWHO}{0.40}
\newcommand{\nMbNNDomMedianBge}{0.00}
\newcommand{\nMbNNBankGAD}{0.86}

\newcommand{\nDAlAucStance}{0.55}
\newcommand{\nDAlAucStanceCI}{0.43 to 0.67}
\newcommand{\nDAlAucCos}{0.66}

\newcommand{\nDAlAucCompass}{0.78}
\newcommand{\nDAlAucCompassCI}{0.68 to 0.88}

\newcommand{\nDAlCorCosTotalTopic}{0.80}

\newcommand{\nDAlCorCompassTopic}{0.53}

\newcommand{\nCAlAucStance}{0.68}
\newcommand{\nCAlAucStanceCI}{0.60 to 0.76}
\newcommand{\nCAlAucCos}{0.56}
\newcommand{\nCAlDiff}{0.117}
\newcommand{\nCAlDiffCI}{\ensuremath{-}0.001 to 0.232}
\newcommand{\nCAlAucCompass}{0.65}
\newcommand{\nCAlAucCompassCI}{0.56 to 0.73}

\newcommand{\nCAlStanceTask}{0.66}

\newcommand{\nCAlStanceBond}{0.65}

\newcommand{\nCAlStanceGoal}{0.72}

\newcommand{\nCAlCorStanceTotalTopic}{\ensuremath{-}0.20}

\newcommand{\nCAlCorCompassLogTurns}{0.45}

\newcommand{\nDMiniHtwoCI}{\ensuremath{-}0.14 to 0.13}

\newcommand{\nDNliShufGeo}{0.33}

\newcommand{\nDNliShufG}{0.49}
\newcommand{\nDNliShufGNull}{\ensuremath{-}0.15 to 0.14}
\newcommand{\nDNliShufPHQ}{0.54}
\newcommand{\nDNliShufPHQNull}{\ensuremath{-}0.15 to 0.14}

\newcommand{\nCMiniHtwoCI}{\ensuremath{-}0.10 to 0.12}

\newcommand{\nCBgeHtwoCI}{\ensuremath{-}0.13 to \ensuremath{-}0.00}

\newcommand{\nCNliHtwoCI}{0.02 to 0.16}

\newcommand{\nCNliShufG}{0.60}

\newcommand{\nCNliShufPHQ}{0.57}

\newcommand{\nCNliNNr}{\ensuremath{-}0.001}

\newcommand{\nNNrMin}{\ensuremath{-}0.045}
\newcommand{\nNNrMax}{0.060}
\newcommand{\nNNRefMin}{0.88}
\newcommand{\nNNRefMax}{0.91}
\newcommand{\nNNpMin}{0.19}

\newcommand{\nDMiniPhqR}{0.15}
\newcommand{\nDMiniPhqPart}{0.07}

\newcommand{\nDBgePhqR}{0.26}
\newcommand{\nDBgePhqPart}{0.16}

\newcommand{\nDNliPhqPart}{0.52}
\newcommand{\nDNliPhqPartCI}{0.43 to 0.61}

\newcommand{\nCMiniPhqR}{0.39}
\newcommand{\nCMiniPhqPart}{0.12}

\newcommand{\nCBgePhqR}{0.53}
\newcommand{\nCBgePhqPart}{0.40}

\newcommand{\nCNliPhqPart}{0.49}
\newcommand{\nCNliPhqPartCI}{0.33 to 0.63}

\newcommand{\nBlockNliCP}{0.031}
\newcommand{\nBlockNliCPHolmish}{0.092}
\newcommand{\nPCLZeroPerson}{683}
\newcommand{\nNoSixEightThreeNliHtwoP}{0.049}
\newcommand{\nHlqcSharedID}{53}

\newcommand{\nHlqcUnresolved}{21}

\newcommand{\nHlqcUnresolvedNeg}{16}
\newcommand{\nHlqcTextDependent}{8}
\newcommand{\nHlqcVideos}{156}
\newcommand{\nTextThreshold}{0.3}
\newcommand{\nGapBelow}{0.027}
\newcommand{\nGapAbove}{0.54}

\newcommand{\nAlTruncShare}{13.9}
\newcommand{\nAlTruncSegments}{2,043}
\newcommand{\nAlSensAucMin}{0.68}
\newcommand{\nAlSensAucMax}{0.69}
\newcommand{\nAlSensLowMin}{\ensuremath{-}0.019}
\newcommand{\nAlSensLowMax}{0.013}
\newcommand{\nEstWhiteFour}{0.51}
\newcommand{\nEstWhiteFourSD}{0.05}
\newcommand{\nInvFailG}{67}
\newcommand{\nInvFailDiag}{153}
\newcommand{\nInvIndef}{200}
\newcommand{\nInvAdmN}{47}
\newcommand{\nInvAdmR}{0.12}
\newcommand{\nInvAdmSD}{0.08}
\newcommand{\nInvPsdR}{0.20}
\newcommand{\nInvPsdSD}{0.11}

\newcommand{\nScMatchNullWord}{12.0}
\newcommand{\nScMatchNullWithin}{4.5}

\newcommand{\nScStrongNullWord}{7.0}

\newcommand{\nScDirNullWithin}{86.5}

\newcommand{\nScWithinNullWithin}{97.5}

\newcommand{\nScWithinPowWordR}{0.27}
\newcommand{\nScWithinPowWithinR}{\ensuremath{-}0.11}
\newcommand{\nCondM}{132}
\newcommand{\nMbDupTexts}{12}
\newcommand{\nSingletonWindows}{13}
\graphicspath{{figures/}}

\newcommand{\PaperTitle}{COMPASS 2.0: psychometric representational similarity analysis distinguishes symptom structure from personal signal}
\title{\PaperTitle}

\author{Baihan Lin$^{1,2,3,4,5}$\\[3pt]
  \parbox{0.92\textwidth}{\centering\normalfont\small
  $^1$Department of Artificial Intelligence and Human Health, Icahn School of Medicine at Mount Sinai, New York, NY, USA\\
  $^2$Department of Psychiatry, Icahn School of Medicine at Mount Sinai, New York, NY, USA\\
  $^3$Department of Neuroscience, Icahn School of Medicine at Mount Sinai, New York, NY, USA\\
  $^4$Mental Illness Research, Education and Clinical Center, James J. Peters VA Medical Center, Bronx, NY, USA\\
  $^5$Berkman Klein Center for Internet \& Society, Harvard University, Cambridge, MA, USA\\[3pt]
  Correspondence: \texttt{baihan.lin@mssm.edu}}}

\renewcommand{\shorttitle}{COMPASS 2.0: psychometric representational similarity analysis}
\renewcommand{\headeright}{Preprint}
\renewcommand{\undertitle}{Preprint}
\hypersetup{pdftitle={\PaperTitle}, pdfauthor={Baihan Lin}}

\date{}
\begin{document}
\maketitle

\begin{abstract}
Language models can score psychiatric questionnaires from speech, but agreement with self-report may reflect the questionnaire rather than the person. We introduce psychometric representational similarity analysis, a framework for comparing the structure of speech-derived scores, self-report, item wording and theory, and implement it alongside person-level construct scoring in COMPASS 2.0. We show how similarly worded items induce covariance without psychological signal. In pre-registered discovery and confirmation analyses of clinical interviews from \nAllPersons{} participants, language-derived symptom geometry resembled wording more than self-report, with no structure beyond wording detected by the registered tests. Geometric agreement with self-report survived assigning participants someone else's answers, whereas person-paired scores captured distress more than specific symptoms. Complementary analyses examined counselling quality and wording structure across \nMbInstruments{} instruments and the Research Domain Criteria (RDoC) framework. These findings distinguish agreement about psychological structure from evidence that language-derived assessments track individual people.
\end{abstract}

\section*{Introduction}

A questionnaire is a list of sentences that becomes, through a person's answers, a map of their mind. When depression items rise and fall together, we infer symptoms that travel together. When alliance items cohere, we infer a working relationship. The sentences give these patterns meaning, but the evidence is meant to come from the person answering them. A century of psychometrics has exposed exceptions, from negatively worded items' method factors\cite{distefano2006method} to scales sharing names or words but not constructs\cite{larsen2016construct,rosenbusch2020semantic}. Item-text similarity alone predicts much of the correlation among survey responses\cite{arnulf2014predicting,rosenbusch2020semantic,wulff2025semantic,hommel2025language,pellert2026values}.

Language-based assessment reverses this arrangement. Rather than asking people to rate items, sentence embeddings, entailment models and large language models score how closely clinical interviews, therapy dialogue or online posts match them\cite{kjell2019semantic,kjell2024beyond,milintsevich2023symptom,lin2023deep,lin2024wat,lin2025compass}. Researchers sum these scores into scales, correlate them into networks and compare groups as though they were questionnaire answers. Yet similarly worded items attract the same speech whatever the speaker feels. Their correlations can reproduce the questionnaire's wording without psychological signal. The resulting map may look coherent, replicate across samples and resemble self-report. Each success invites the same question: how much did the person contribute to the structure we recognize? A model can inherit a convincing map of symptoms before it has located anyone on that map.

Distinguishing these possibilities requires comparing representations across questionnaire responses, clinicians' ratings, model outputs and theory. Representational similarity analysis addressed an analogous problem in neuroscience by expressing brain activity, behaviour and computational models as dissimilarity matrices over shared conditions\cite{kriegeskorte2008rsa,diedrichsen2017representational,lin2024topology}. We introduce psychometric representational similarity analysis: to our knowledge, the first framework to compare language-derived questionnaire representations with self-report, wording and theory within this logic. Instrument items become the conditions. Each source supplies an item-by-item geometry: co-endorsement in self-report, co-activation by speech, similarity of item meanings or theoretical grouping. The source can also be another speaker, session or time point. These psychometric representational dissimilarity matrices (RDMs) summarize how each source organizes constructs and permit direct comparison (Fig.~\ref{fig:hub}). Geometry describes a population or one speaker's language over time. Whether a measure tracks the right person requires a separate, person-paired test.

Clinical applications also require scores. COMPASS (Computational Mapping of Psychometric Assessment in Speech and Semantics) applies instrument scoring keys to the same item activity profiles, estimating symptoms, alliance or other constructs, following them over a conversation and comparing them between speakers. Our earlier version introduced alliance scoring in psychotherapy transcripts\cite{lin2025compass}, building on item-based dialogue annotation\cite{lin2023deep,lin2024wat}, and we subsequently applied it to personality\cite{lin2025personality}. COMPASS 2.0 connects representations and construct scores, allowing each score to be examined against its underlying geometry (Fig.~\ref{fig:method}).

Both layers inherit the same confound. Scoring speech embedding $\mathbf{e}$ against item embeddings $A$ gives activities $\mathbf{p}=A^{\top}\mathbf{e}$ with covariance $A^{\top}\Sigma A$. Under isotropic speech variation ($\Sigma\propto I$), their correlation geometry is exactly the item similarity $A^{\top}A$. This is the wording null: the geometry expected from isotropic speech carrying no information about the person. Consider the near-duplicate PHQ-8 and PCL-C sleep items. Sleep-related speech raises both scores whether or not depressive and post-traumatic sleep problems co-occur. Because related constructs are deliberately described in related words, this inherited geometry can look psychologically valid. If wording also shapes self-report, the two measurements can agree through a structure that both inherit. Agreement then needs an explanation of its source before it can support a claim about the person. Neuroscience likewise had to account for stimulus and noise structure before interpreting neural geometry\cite{walther2016reliability,diedrichsen2021wuc}. Topic, length and interviewer also shape speech, so isotropy is a reference, not a complete account of nonpsychological variation. COMPASS 2.0 whitens against wording, treating departures as diagnostic rather than necessarily personal. It checks an empirical within-speaker reference, appropriate only when that variation is nuisance, centres activities on the interview script, tests residual agreement with self-report by item permutation, and scores stance alongside topic.

We tested this framework in pre-registered analyses of two clinical interview corpora with item-level self-reports, re-examined earlier alliance scores in psychotherapy and counselling, and compared wording across \nMbInstruments{} instruments and the Research Domain Criteria\cite{insel2010rdoc}. In both interview samples and both embedding models, geometry followed wording more closely than self-report. No scorer showed registered evidence of population structure beyond wording. A small exploratory entailment effect disappeared under another reference. Geometric agreement ignored who said what, whereas paired scores tracked distress, for entailment beyond interview length and topic, with a small item-specific excess for embedding scorers in confirmation. Earlier keyed Bond scores tracked relationship talk. Stance scores distinguished counselling quality in a second corpus without robustly outperforming keyed scores of the same items. Across instruments, wording clustered by construct domain, making these controls relevant wherever item texts probe language.

\section*{Results}

\subsection*{Psychometric RDMs place language, self-report, wording and theory in one space}

\begin{figure}[!tbp]
  \centering
  \includegraphics[width=\textwidth]{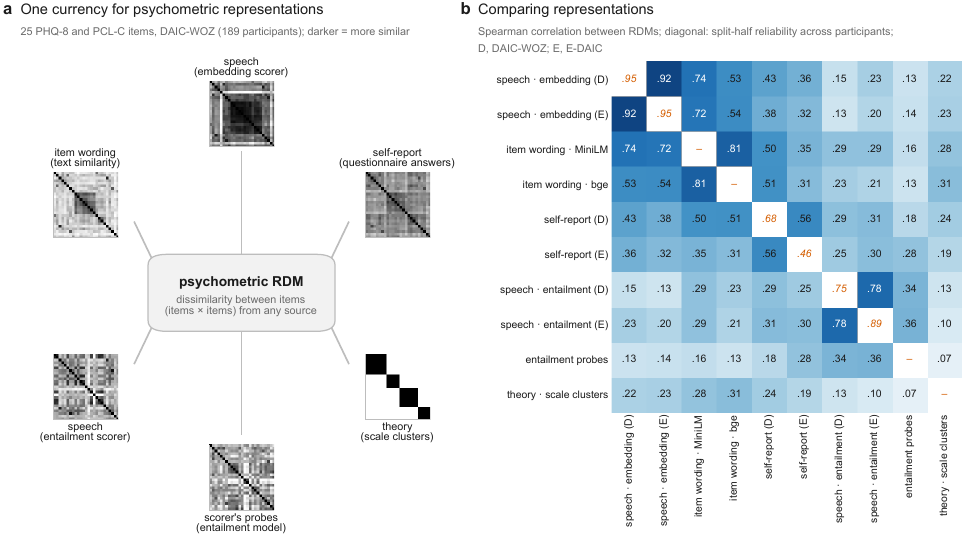}
  \caption{\textbf{Psychometric representational dissimilarity matrices give every source of psychometric information one format.} \textbf{a}, For the same 25 items (8 PHQ-8, 17 PCL-C), each source yields an item-by-item RDM ($1-$ correlation or $1-$ similarity; darker, more similar; DAIC-WOZ, 189 participants): speech scored by an embedding model or by an entailment model (participants' mean profiles), the participants' self-reports, the similarity of the item texts, the entailment scorer's confusions between item probes, and a theory model in which items of the same scale cluster (PHQ-8; PCL-C re-experiencing, avoidance and numbing, hyperarousal) are similar. \textbf{b}, Spearman correlations between the RDMs' entries over item pairs, for sources in DAIC-WOZ (D) and E-DAIC (E). Diagonal: split-half reliability of each sampled RDM across random halves of participants (Spearman--Brown corrected; 200 splits); the attainable correlation between two sources is about the square root of the product of their reliabilities; dashes mark sources without sampling noise.}
  \label{fig:hub}
\end{figure}

Psychometric representational similarity analysis asks each source how the items relate. Similarity means co-endorsement across people for self-report, co-activation by speech for language, related meanings for wording, and shared instrument clusters for theory. Each yields an RDM, $D_{jk}=1-r_{jk}$ over items $j$ and $k$. Correlating its item-pair entries with another RDM compares the sources (Fig.~\ref{fig:hub}a).

For the 25 PHQ-8 and PCL-C items, these comparisons reveal the central pattern (Fig.~\ref{fig:hub}b). Embedding-scored speech geometry was nearly identical across samples ($\rho=\nHubEmbDE{}$) and closer to wording (\nHubEmbWordD{}, \nHubEmbWordE{}) than self-report (\nHubEmbSelfD{}, \nHubEmbSelfE{}). Self-report agreed across samples (\nHubSelfDE{}) and partly with wording (\nHubSelfWordD{}, \nHubSelfWordE{}). Entailment geometry followed wording less (\nHubEntWordD{}) but agreed across samples (\nHubEntDE{}) and partly with its own probe confusions (\nHubEntProbeD{}, \nHubEntProbeE{}). The symptom-cluster theory model related weakly to every source ($\rho\le\nHubTheoryMax{}$). Under classical assumptions, split-half reliabilities $r_1$ and $r_2$ across random participant halves bound attainable agreement at about $\sqrt{r_1r_2}$. Reliability was \nRelEmbD{} for embedding-scored speech and \nRelSelfD{} and \nRelSelfE{} for self-report (diagonal). Self-reports agreed across samples as well as these bounds allowed. Language reached about half its bound with self-report (\nHubEmbSelfDNorm{} after correction) and three quarters with wording (\nHubEmbWordDNorm{}). The following tests examine these relationships formally. The high reproducibility of the language geometry makes its source especially consequential: a structure can be stable across populations because the measurement procedure carries it from one population to the next.

\begin{figure}[!t]
  \centering
  \includegraphics[width=\textwidth]{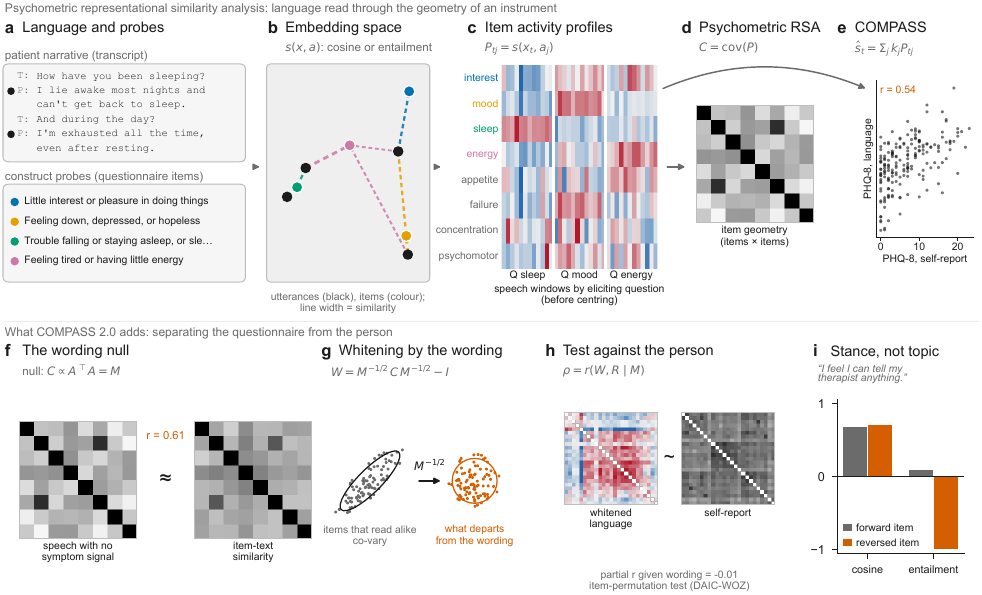}
  \caption{\textbf{COMPASS 2.0 reads speech through an instrument as representations (psychometric RSA) and as construct scores (COMPASS), with controls for what the probes impose.} \textbf{a}, The two inputs: language (invented transcript lines) and the items of an instrument (four PHQ-8 items). \textbf{b}, Both are embedded in one space (multidimensional scaling of MiniLM embeddings of the example sentences); lines join utterances to the items they resemble, with width proportional to cosine similarity. \textbf{c}, Item activity profiles: the similarity of each speech window to each of the eight PHQ-8 items (z-scored per item), for simulated windows with no symptom signal, grouped by the question that elicited them, before centring. \textbf{d}, Psychometric RSA: the item geometry, here the correlations between the items' activities across the windows in \textbf{c} after centring each window on its question's mean. \textbf{e}, COMPASS: keyed construct scores; here a language-derived PHQ-8 score (entailment scorer, participants' mean profiles) against the self-reported PHQ-8 total (DAIC-WOZ, 189 participants). \textbf{f}, The wording null: the geometry of speech without symptom signal resembles the similarity of the item texts. \textbf{g}, Whitening by the item-text similarity matrix $M$ factors out the covariance expected from the item texts alone. \textbf{h}, The whitened language geometry is compared with the self-reported item geometry while controlling for wording (DAIC-WOZ, 25 PHQ-8 and PCL-C items, MiniLM; diagonal omitted). \textbf{i}, Cosine similarity and entailment-model stance, $P(\text{entailment})-P(\text{contradiction})$, of an example utterance with a forward alliance statement and its reversal (paraphrases, not inventory items).}
  \label{fig:method}
\end{figure}

\subsection*{From speech to psychometric representations and construct scores}

Figure~\ref{fig:method} follows speech through both layers of COMPASS 2.0. We divide speech into units $x_1,\dots,x_T$ (answer windows, turns or short exchanges) and place them in a shared space with the items $a_1,\dots,a_N$ of an instrument (Fig.~\ref{fig:method}a,b). Every unit is scored against every item,
\begin{equation}
P_{tj}=s(x_t,a_j),
\label{eq:activity}
\end{equation}
where $s$ is the cosine similarity of their embeddings or an entailment model's stance, so that each row of $P$ is an item activity profile (Fig.~\ref{fig:method}c). Psychometric representational similarity analysis summarizes $P$ by its item-by-item geometry, the covariance
\begin{equation}
C=\tfrac{1}{T-1}\,\bigl(P-\mathbf{1}\bar{\mathbf{p}}^{\top}\bigr)^{\top}\bigl(P-\mathbf{1}\bar{\mathbf{p}}^{\top}\bigr),
\label{eq:geometry}
\end{equation}
or the corresponding correlations $r_{jk}$ and RDM $D_{jk}=1-r_{jk}$, computed across the units of one person or across people's mean profiles (Fig.~\ref{fig:method}d). COMPASS reads the same profiles as construct scores, weighting them by an instrument's scoring key $\mathbf{k}$,
\begin{equation}
\hat{s}_t=\textstyle\sum_j k_jP_{tj},
\label{eq:score}
\end{equation}
which can be followed over a conversation, compared between speakers or averaged per person (Fig.~\ref{fig:method}e). In DAIC-WOZ, for example, a language-derived PHQ-8 score from the entailment scorer correlated \nMethodRScore{} with the participants' self-reported PHQ-8 totals.

The wording null enters both read-outs. COMPASS 2.0 addresses it in four steps (Fig.~\ref{fig:method}f--i). With cosine scoring, $P=EA$ for unit embeddings $E$ and item embeddings $A$, so that $C=A^{\top}\Sigma_EA$; when speech carries no item-specific signal ($\Sigma_E\propto I$), the geometry reduces to the item-text similarity $M=A^{\top}A$ (Fig.~\ref{fig:method}f; $r=\nMethodRWording{}$ in the illustration). Whitening factors this prediction out,
\begin{equation}
W=M^{-1/2}\,C\,M^{-1/2}-I,
\label{eq:whiten}
\end{equation}
with off-diagonal entries that vanish under the null (Fig.~\ref{fig:method}g). Under signal, $W$ is a transformed image of the person's structure rather than its recovery, so we use it as a diagnostic of departure from the reference. The within-speaker covariance of these activities offers an empirical alternative to $M$, capturing topic and other structure within each conversation. It is a valid reference only if between-speaker nuisance has the same shape as within-speaker variation and the latter carries no symptom structure, so we use it as a sensitivity analysis rather than as an identification of where an effect comes from (Methods). The remainder is compared with the self-reported item correlations $R^{\mathrm{self}}$ beyond the wording, by the partial correlation over item pairs
\begin{equation}
\rho=r\bigl(W,R^{\mathrm{self}}\mid M\bigr),
\label{eq:test}
\end{equation}
with item-permutation inference (Fig.~\ref{fig:method}h); in interviews, each unit is first centred on the mean activity of the question that elicited it, computed from other people. Keyed scores also require stance sensitivity: cosine similarity barely distinguishes a statement from its reversal. For the example utterance in Fig.~\ref{fig:method}i, cosine similarity was \nStanceCosFwd{} with a forward alliance statement and \nStanceCosRev{} with its reversal, whereas the entailment model scored them \nStanceNliFwd{} and \nStanceNliRev{}.

\subsection*{Item wording fixes the geometry of language-derived scores}

\begin{figure}[!tbp]
  \centering
  \includegraphics[width=\textwidth]{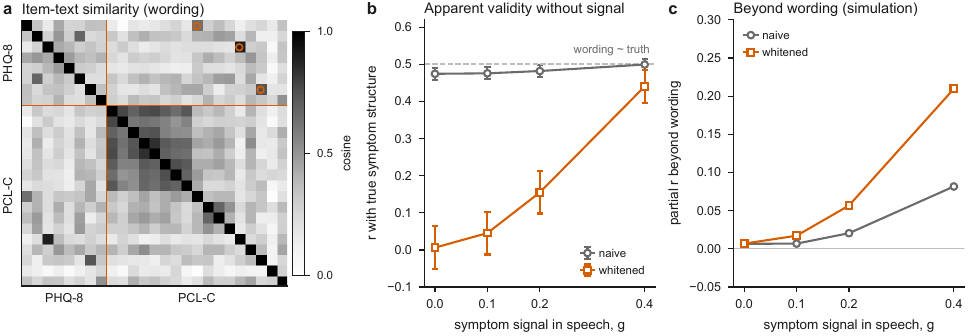}
  \caption{\textbf{Item wording alone produces a convincing symptom structure.} \textbf{a}, Cosine similarity between the texts of the 8 PHQ-8 and 17 PCL-C items (MiniLM). Circles mark near-duplicate items across the two instruments (sleep, concentration, loss of interest). \textbf{b}, Simulation: correlation of the language-derived item geometry with the true symptom structure as the symptom signal in speech grows (mean $\pm$ s.d.\ over 200 simulated samples of 189 people). Without any signal ($g=0$), the naive geometry already correlates almost as strongly with the truth as the wording itself does (dashed line), because the truth partly resembles the wording. The whitened geometry starts at zero. \textbf{c}, Partial correlation with the truth beyond the part predicted by wording.}
  \label{fig:null}
\end{figure}

We first measured the structure present in the probes before anyone speaks. The similarities among the 8 PHQ-8 and 17 PCL-C items\cite{kroenke2009phq8,blanchard1996pcl} concentrate in about \nWordRankMini{} effective dimensions in MiniLM sentence embeddings\cite{reimers2019sbert,wang2020minilm} and \nWordRankBge{} in bge-large embeddings\cite{xiao2024cpack} (participation ratio of the item-similarity matrix\cite{gao2017dimensionality}; the matrices are full rank, with condition number \nCondM{} for MiniLM), and three pairs of items across the two instruments are near-duplicates: sleep (cosine \nDupSleep{}), concentration (\nDupConc{}) and loss of interest (\nDupInterest{}; Fig.~\ref{fig:null}a).

We simulated people whose true item correlations shared half their structure with wording, as survey correlations do\cite{arnulf2014predicting,hommel2025language}, and varied expression of their item levels in speech by strength $g$ (Methods). With no symptom signal ($g=0$), naive geometry correlated \nSimNaiveZero{} with true symptom structure entirely through wording (Fig.~\ref{fig:null}b). Whitening ($M^{-1/2}CM^{-1/2}-I$) removed this apparent validity ($r=\nSimWhiteZero{}$) and increasingly tracked true structure as signal grew ($r=\nSimWhiteFour{}$ at $g=0.4$). Beyond wording-predicted truth, whitened geometry outperformed naive geometry (partial $r=\nSimWhiteBeyondFour{}$ versus \nSimNaiveBeyondFour{}; Fig.~\ref{fig:null}c). Registered Freedman--Lane item-permutation inference\cite{freedman1983nonstochastic,winkler2014permutation} remained near its nominal false-positive rate (\nFLNullN{}\% of \nFLReps{} samples of 189 people, \nFLNullS{}\% of samples of 86, $\alpha=0.05$). Power at 189 people was \nFLPowerFourN{}\% for strong structure ($g=0.4$) but only \nFLPowerTwoN{}\% for moderate structure ($g=0.2$). 

Exact deconvolution recovers population structure when $g>0$ and its diagonal fit is identifiable, but often failed in samples. At $g=0.4$, fitted signal variance was non-positive in \nInvFailG{} of 200 samples, a noise-subtracted variance was non-positive in \nInvFailDiag{}, and all \nInvIndef{} noise-subtracted matrices were indefinite. Correlation with truth was \nInvAdmR{} $\pm$ \nInvAdmSD{} among the \nInvAdmN{} admissible fits, a selected subset, and \nInvPsdR{} $\pm$ \nInvPsdSD{} after positive-semidefinite projection of all fits, compared with \nEstWhiteFour{} $\pm$ \nEstWhiteFourSD{} for whitening (Extended Data Fig.~\ref{fig:edsim}a). These compare structural agreement, not estimators of a common target: whitening does not recover true structure. 

Reference calibration depended on speech variation (Extended Data Fig.~\ref{fig:edsim}c). With identically shaped topic variation between and within speakers, wording whitening was liberal (\nScMatchNullWord{}\% false positives without person signal), whereas within-speaker whitening was not (\nScMatchNullWithin{}\%). The within-speaker reference failed when variation followed different directions or contained genuine symptom dynamics (\nScDirNullWithin{}\% and \nScWithinNullWithin{}\% false positives, respectively, the latter with spurious negative associations). Wording-reference rates remained between \nScStrongNullWord{}\% and \nScMatchNullWord{}\%. With genuine within-speaker dynamics, within-speaker whitening also removed real person structure and reversed its sign at $g=0.4$ (mean partial $r=\nScWithinPowWithinR{}$, versus \nScWithinPowWordR{} under $M$).

\begin{figure}[!t]
  \centering
  \includegraphics[width=\textwidth]{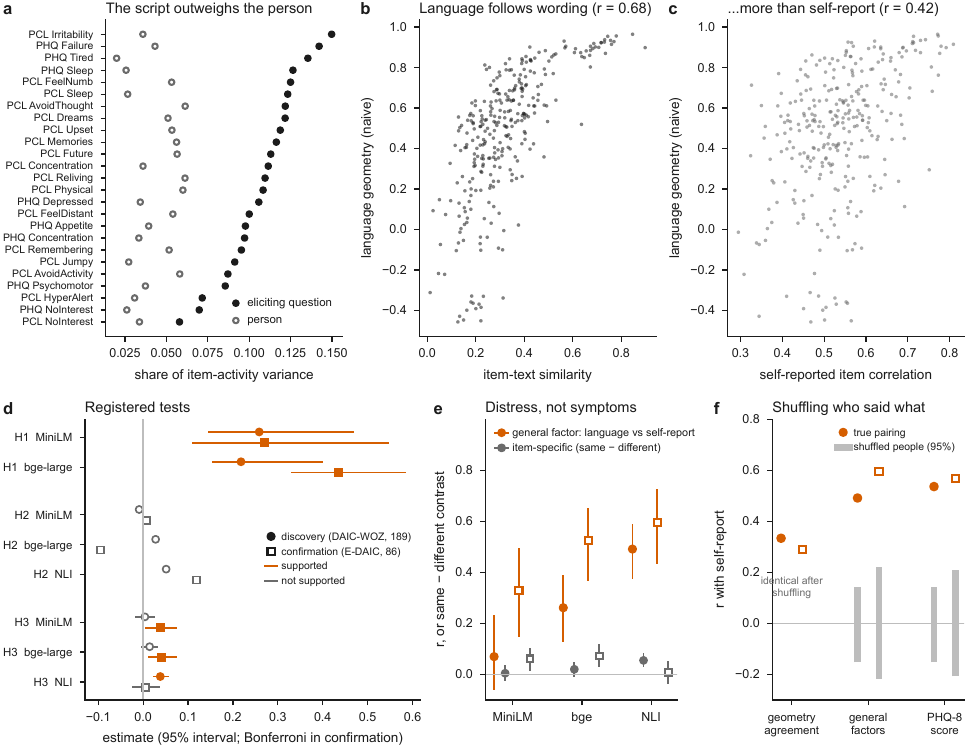}
  \caption{\textbf{In clinical interviews, language-derived symptom structure follows the questionnaire and the script, and geometric agreement is blind to who said what.} \textbf{a}, Share of the variance of each item's activity across answer windows explained by the eliciting question (filled) and by the person (open); DAIC-WOZ, MiniLM. \textbf{b}, Item-pair correlations of the naive language-derived geometry (189 participants) against item-text similarity. \textbf{c}, The same correlations against the participants' self-reported item correlations. \textbf{d}, The registered tests in discovery (DAIC-WOZ; circles, 95\% intervals) and confirmation (E-DAIC; squares, Bonferroni-adjusted intervals). H1 compares the naive geometry's correlation with wording and with self-report; H2 is the partial correlation of the whitened geometry with self-report beyond wording (no interval; Freedman--Lane test); H3 is the mean same-item minus different-item correlation between language and self-report. Accent marks a supported test. \textbf{e}, Exploratory: correlation between the general factors (first principal components) of language-derived and self-reported items (accent), and item-specific convergence after removing the self-reported general factor (grey; a same-minus-different contrast, not a variance share), with person-bootstrap 95\% intervals; circles discovery, squares confirmation. \textbf{f}, Person shuffle (entailment scorer): agreement of the naive language geometry with the self-reported geometry, of the general factors, and of a COMPASS PHQ-8 score with the self-reported total, with true pairing (accent; circles discovery, squares confirmation) and after shuffling which self-report belongs to which participant (grey, 95\% range of 1,000 shuffles).}
  \label{fig:interviews}
\end{figure}

\subsection*{Interview scripts impose a second structure}

Interview questions impose their own structure. In DAIC-WOZ's scripted virtual interviews\cite{gratch2014daic,devault2014simsensei}, question identity was associated with a median \nDaicEtaQ{} of item-activity variance across answer windows (one-way $\eta^2$, range \nDaicEtaQlo{}--\nDaicEtaQhi{}), compared with \nDaicEtaP{} for person identity (Fig.~\ref{fig:interviews}a). In E-DAIC\cite{ringeval2019avec}, interview phase accounted for \nEdaicEtaQ{} and person identity for \nEdaicEtaP{}. Answering different question subsets therefore shapes participants' averages. Before averaging, we centred each window on its question's mean among other participants (Methods). This removes estimated question means, not every script effect.

Geometry correlated \nDaicRWindow{} with wording across windows and \nDaicRPerson{} across people (Fig.~\ref{fig:interviews}b), falling to \nDaicRPersonCentred{} after centring and \nDaicRWhitened{} after whitening. Whitened population geometry was reliable (split-half reliability \nDaicRelWhite{} in DAIC-WOZ, \nEdaicRelWhite{} in E-DAIC). Individual deviations were not (median \nDaicRelDev{}, \nEdaicRelDev{}; Extended Data Fig.~\ref{fig:edreliab}). One interview supplied too little language for individual item geometry, so our tests concern population structure.

\subsection*{Registered tests separate the wording from the symptoms}

We registered three tests before reading any self-reported item score (Methods). Discovery used DAIC-WOZ (\nDaicPersons{} participants), and confirmation used E-DAIC (\nEdaicPersons{} participants, autonomous interviews and automatic transcripts). Scorers were MiniLM and bge cosine similarity, and an entailment model\cite{williams2018mnli,he2023debertav3,yin2019zeroshot} assessing support or contradiction of first-person symptom statements.

H1 tested whether naive geometry resembled wording more than self-reported item correlations. It did in discovery (MiniLM: $r=\nDMiniHoneW{}$ versus \nDMiniHoneS{}, difference \nDMiniHoneD{}, 95\% confidence interval (CI) \nDMiniHoneCI{}; bge: \nDBgeHoneW{} versus \nDBgeHoneS{}, difference \nDBgeHoneD{}, CI \nDBgeHoneCI{}) and confirmation (MiniLM: \nCMiniHoneW{} versus \nCMiniHoneS{}, difference \nCMiniHoneD{}, 97.5\% CI \nCMiniHoneCI{}; bge: \nCBgeHoneW{} versus \nCBgeHoneS{}, difference \nCBgeHoneD{}, CI \nCBgeHoneCI{}; Fig.~\ref{fig:interviews}b--d). Self-reported correlations themselves followed wording\cite{arnulf2014predicting,hommel2025language} (MiniLM: \nDMiniSrWord{} discovery, \nCMiniSrWord{} confirmation). Shared wording thus accounted for much of naive geometry's apparent validity. Exploratory adjustment for wording reduced agreement with self-report from \nDMiniNaiveSr{} to \nDMiniNaiveSrPartial{} in discovery and \nCMiniNaiveSr{} to \nCMiniNaiveSrPartial{} in confirmation (bge: \nDBgeNaiveSr{} to \nDBgeNaiveSrPartial{}, and \nCBgeNaiveSr{} to \nCBgeNaiveSrPartial{}).

The primary test, H2, asked whether centred, whitened language geometry matched self-report beyond wording. It was not supported. Partial correlations were \nDMiniHtwoR{}, \nDBgeHtwoR{} and \nDNliHtwoR{} in discovery (MiniLM, bge, entailment; Freedman--Lane $P=\nDMiniHtwoP{}$, \nDBgeHtwoP{}, \nDNliHtwoP{}) and \nCMiniHtwoR{}, \nCBgeHtwoR{} and \nCNliHtwoR{} in confirmation (Holm-adjusted $P=\nCMiniHtwoPholm{}$, \nCBgeHtwoPholm{}, \nCNliHtwoPholm{}). Exploratory person-bootstrap intervals admitted small effects in either direction, including MiniLM's \nDMiniHtwoCI{} and \nCMiniHtwoCI{}. In confirmation, entailment's interval excluded zero (\nCNliHtwoCI{}) and bge's lay below zero (\nCBgeHtwoCI{}). Permuting items only within instruments gave $P=\nBlockNliCP{}$ for entailment using the same registered statistic (unadjusted, approximately \nBlockNliCPHolmish{} after Holm adjustment across scorers). Simulation power was high only for strong structure (\nFLPowerFourN{}\% at $g=0.4$, \nFLPowerTwoN{}\% at $g=0.2$), leaving moderate structure beyond wording unresolved. 

The residual also depended on the reference. Within-speaker covariance closely resembled wording for embedding scorers ($r=\nNNRefMin{}$ to \nNNRefMax{}). Whitening against it with registered controls unchanged gave partial correlations from \nNNrMin{} to \nNNrMax{} (all $P\ge\nNNpMin{}$; Supplementary Table~\ref{tab:sensitivity}), including \nCNliNNr{} for confirmation entailment. Its validity depends on assumptions the data cannot verify (Extended Data Fig.~\ref{fig:edsim}c), so this establishes reference sensitivity, not a nuisance origin. Excluding one confirmation participant with a PCL-C response of 0, outside the 1--5 scale, left every registered conclusion unchanged.

H3 tested whether language-derived scores correlated more with the matching self-reported item than with other items. Specificity was small and inconsistent. Entailment supported H3 in discovery (difference \nDNliHthreeD{}, CI \nDNliHthreeCI{}) but not confirmation (\nCNliHthreeD{}, 98.3\% CI \nCNliHthreeCI{}). Both cosine scorers supported it in confirmation (MiniLM \nCMiniHthreeD{}, CI \nCMiniHthreeCI{}; bge \nCBgeHthreeD{}, CI \nCBgeHthreeCI{}) but not discovery (Fig.~\ref{fig:interviews}d; Extended Data Table~\ref{tab:registered}).

H1 and H2 compare population geometries, which ignore person pairing. H3 and the following read-outs retain it. Reassigning self-reports between participants left geometric agreement exactly unchanged (discovery entailment: \nDNliShufGeo{} before and after), while paired measures fell to their null (Fig.~\ref{fig:interviews}f). General-factor agreement fell from \nDNliShufG{} to a shuffled range of \nDNliShufGNull{}, and COMPASS PHQ-8 agreement with self-reported totals from \nDNliShufPHQ{} to \nDNliShufPHQNull{} (confirmation before shuffling: \nCNliShufG{}, \nCNliShufPHQ{}). This invariance follows from the comparison itself. Reordering the people leaves the correlations among questionnaire items unchanged, so the population RDM retains no record of which report belongs to which interview. Geometry can establish that language organizes items as self-report does. Establishing that it measures the right person requires paired read-outs.

\subsection*{Speech conveys how distressed a person is more than which symptoms they have}

What does speech convey if symptom profiles are poorly recovered? Exploratory analyses compared the first principal components of language-derived and self-reported items from both scales across participants, then examined item-specific correspondence after removing the self-reported general factor (Methods). These components are sample-specific summaries, not independently validated general factors. For entailment, they correlated $r=\nDNliG{}$ (95\% CI \nDNliGCI{}) in discovery and \nCNliG{} (\nCNliGCI{}) in confirmation. Mean language-derived item scores correlated \nDNliMeanPHQ{} and \nCNliMeanPHQ{} with PHQ-8 totals. Cosine scorers carried less general-factor information (bge \nDBgeG{}, \nCBgeG{}; MiniLM \nDMiniG{}, \nCMiniG{}). Entailment's association persisted beyond speech length and questionnaire-topic activation: after partialling answer-window count, word count and topic activation, COMPASS PHQ-8 scores correlated \nDNliPhqPart{} (95\% CI \nDNliPhqPartCI{}) and \nCNliPhqPart{} (\nCNliPhqPartCI{}) with self-reported PHQ-8 totals. Cosine-score correlations shrank (MiniLM: \nDMiniPhqR{}, \nCMiniPhqR{} to \nDMiniPhqPart{}, \nCMiniPhqPart{}; bge: \nDBgePhqR{}, \nCBgePhqR{} to \nDBgePhqPart{}, \nCBgePhqPart{}; Supplementary Table~\ref{tab:baselines}). Item-specific convergence was much smaller (same-item minus different-item correlation \nDMiniSpec{} to \nCBgeSpec{}; Fig.~\ref{fig:interviews}e). Agreement lay mainly in general distress, reminiscent of hierarchical psychopathology models\cite{caspi2014pfactor,kotov2017hitop}, while apparent item structure largely belonged to the questionnaire. The differently scaled statistics do not partition symptom information. Both paired read-outs vanished when self-reports were shuffled (Fig.~\ref{fig:interviews}f). Speech therefore carried information tied to the speaker even where its apparent symptom structure largely reflected the probes. The distinction between these findings is what permits a useful signal to survive a critical test of its interpretation.

\begin{figure}[!t]
  \centering
  \includegraphics[width=\textwidth]{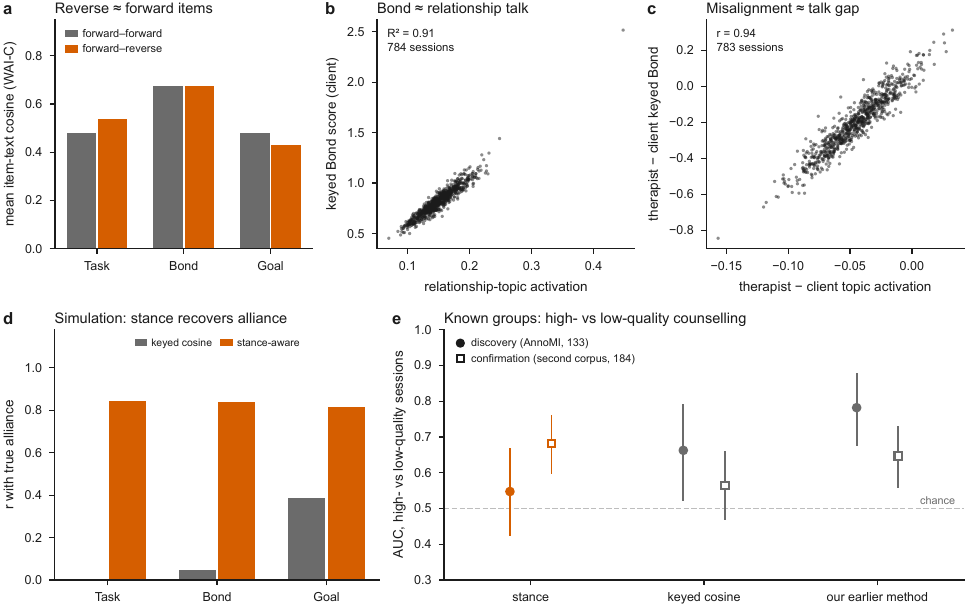}
  \caption{\textbf{Keyed similarity scores of the alliance track what a session talks about, and stance-aware scores separate session quality in a second corpus.} \textbf{a}, Mean cosine similarity between Working Alliance Inventory client-form items with the same key (forward--forward) and with opposite keys (forward--reverse), per subscale (MiniLM). \textbf{b}, Keyed Bond score of client turns (our earlier scoring rule) against the mean similarity of the same turns to all Bond items regardless of key, per session. \textbf{c}, Therapist-minus-client difference in keyed Bond score against the difference in relationship-topic activation. \textbf{d}, Simulation: correlation with the true alliance of keyed cosine scores and of stance-aware scores. \textbf{e}, Known groups: area under the curve (AUC) for separating high- from low-quality counselling sessions by the stance score, by the keyed cosine score of the same 36 observer-form items and by a replica of our earlier scoring, in discovery (AnnoMI; circles) and confirmation (second corpus; squares), with stratified-bootstrap 95\% intervals.}
  \label{fig:alliance}
\end{figure}

\subsection*{Keyed alliance scores track topic, and stance-aware scores separate session quality in a second corpus}

The therapeutic alliance, psychotherapy's most robust process predictor of outcome\cite{fluckiger2018alliance}, raises the same problem. Of the Working Alliance Inventory's 36 items\cite{horvath1989wai,tracey1989wai,hatcher2006waisr}, \nWaiReverse{} are reverse-keyed: relationship discomfort, for example, counts against alliance. COMPASS originally scored client and therapist turns by keyed sums of cosine similarities to their respective forms\cite{lin2025compass}. Yet cosine similarity scarcely distinguishes a statement from its reversal\cite{ettinger2020bert}. Forward and reverse client-form items were as similar in MiniLM (mean \nWAICFR{}) as forward items among themselves (\nWAICFF{}; Fig.~\ref{fig:alliance}a). Keyed cosine sums therefore contrast forward and reverse topics rather than relationship quality. Talking about trust and expressing trust are different acts. A discussion of its absence may activate the same topic as an expression of confidence in the therapist, even though their implications for the alliance point in opposite directions.

Co-activation of forward and reverse items made the keyed Bond score nearly proportional to relationship-topic activation across \nCompSessions{} psychotherapy sessions ($R^2=\nCompClBondRsq{}$ for clients, \nCompThBondRsq{} for therapists; Fig.~\ref{fig:alliance}b). Shared item similarities partly enforce this relation arithmetically, but opposing activation, as stance sensitivity requires, would disrupt it. Previously reported therapist--client alliance misalignment tracked their difference in relationship talk ($r=\nCompGapBond{}$, \nCompBoth{} sessions; Fig.~\ref{fig:alliance}c). Task and Goal scores related little to topic ($R^2\le\nCompClTaskRsq{}$) or clinical collection ($\eta^2\le\nCompEtaCollMax{}$). Simulated keyed cosine Task and Bond scores were uncorrelated with true alliance ($r=\nSimCosTask{}$, \nSimCosBond{}) but followed topic share ($r=\nSimTalkTask{}$, \nSimTalkBond{}). Stance scoring recovered alliance on every subscale ($r\ge\nSimStanceGoal{}$; Fig.~\ref{fig:alliance}d).

We pre-specified a known-groups test, fixing code before comparing any quality label with model output: stance scores of the 36 observer-form items should distinguish counselling quality and outperform keyed cosines of those items. Discovery used AnnoMI\cite{wu2023annomi} (\nAnnomiSessions{} sessions, \nAnnomiHigh{} high and \nAnnomiLow{} low quality). Confirmation used \nHlqcIndep{} sessions from a second corpus\cite{perezrosas2019counselor} (\nHlqcIndepHigh{} high, \nHlqcIndepLow{} low), excluding shared AnnoMI video identifiers (Methods). In discovery, stance scores failed to separate quality groups (area under the curve (AUC) \nDAlAucStance{}, 95\% CI \nDAlAucStanceCI{}). Keyed cosines (AUC \nDAlAucCos{}) and replicated earlier scores (\nDAlAucCompass{}, 95\% CI \nDAlAucCompassCI{}) succeeded, while tracking topic activation ($r=\nDAlCorCosTotalTopic{}$, \nDAlCorCompassTopic{}). 

Confirmation, designated decisive, reversed this ordering (Fig.~\ref{fig:alliance}e). Stance scores separated quality groups (AUC \nCAlAucStance{}, 95\% CI \nCAlAucStanceCI{}), supporting the first test, on every subscale (Task \nCAlStanceTask{}, Bond \nCAlStanceBond{}, Goal \nCAlStanceGoal{}), with only weak negative topic association ($r=\nCAlCorStanceTotalTopic{}$). Superiority over keyed cosines (AUC \nCAlAucCos{}, difference \nCAlDiff{}, 95\% CI \nCAlDiffCI{}) narrowly missed the pre-specified criterion. 

\begin{figure}[!t]
  \centering
  \includegraphics[width=\textwidth]{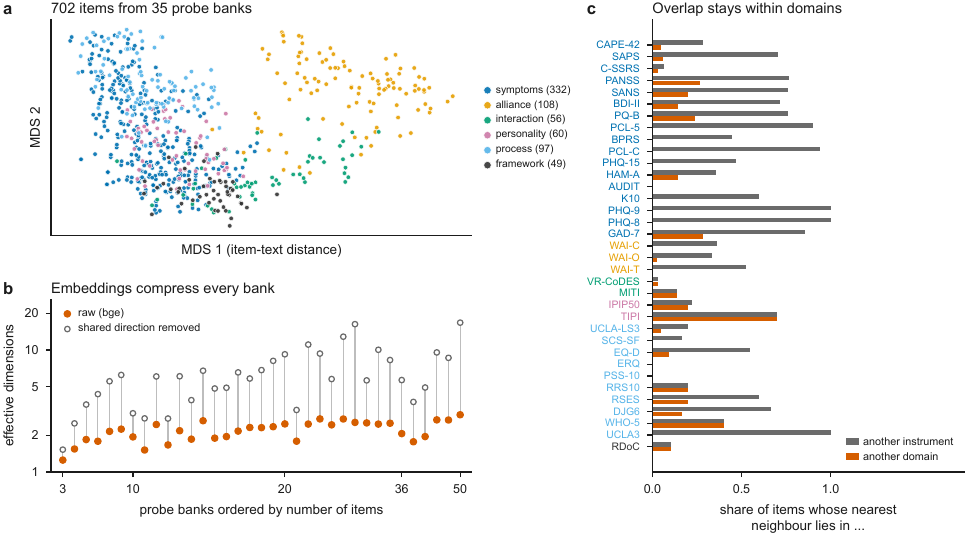}
  \caption{\textbf{Item wording carries a low-dimensional, domain-structured geometry in every instrument examined.} \textbf{a}, Classical multidimensional scaling of the cosine distances between \nMbItems{} item texts and construct definitions from \nMbBanks{} probe banks (MiniLM), coloured by construct domain. \textbf{b}, Effective number of dimensions (participation ratio) of each bank's item-similarity matrix, before (filled) and after (open) removing the direction shared by all item embeddings (bge; banks ordered by number of items; logarithmic scale). \textbf{c}, Share of each bank's items whose nearest neighbour lies in another instrument (grey) or in another construct domain (accent; MiniLM).}
  \label{fig:banks}
\end{figure}

Confirmation contained \nHlqcVideos{} videos, and transcript screening identified \nHlqcTextDependent{} sessions containing AnnoMI material under different identifiers. The corpora were therefore not fully separate. Sensitivity analyses resampled whole videos, removed matched sessions, added \nHlqcUnresolvedNeg{} previously excluded sessions with neither identifier nor transcript matches, or dropped overlong segments (Supplementary Tables~\ref{tab:alliancesens}, \ref{tab:provenance}). Stance AUCs remained \nAlSensAucMin{}--\nAlSensAucMax{}, but the lower confidence bound for superiority ranged from \nAlSensLowMin{} to \nAlSensLowMax{}. Quality separation was robust, superiority was not. Negative screening cannot establish independence of actors, therapists or productions. Replicated earlier scoring also exceeded chance in confirmation (AUC \nCAlAucCompass{}, 95\% CI \nCAlAucCompassCI{}), partly through session length ($r=\nCAlCorCompassLogTurns{}$), and tracked topic and length more closely than stance. Which signal better distinguished quality thus depended on the corpus. Quality labels are not alliance ratings: these results support counselling-quality discrimination, not validation of Task, Bond and Goal scores.

\subsection*{Wording clusters by construct domain across instruments}

We then tested whether this artefact extends beyond the symptom and alliance instruments. We assembled \nMbItems{} items and definitions from \nMbBanks{} probe banks (Methods). They covered clinical symptoms of depression, anxiety, trauma, suicidality, psychosis, somatic distress and alcohol use, rated by patients or by clinicians\cite{kroenke2001phq9,beck1996bdi,kessler2002k10,spitzer2006gad7,hamilton1959hama,blevins2015pcl5,posner2011cssrs,loewy2011pqb,stefanis2002cape,overall1962bprs,kay1987panss,andreasen1982sans,kroenke2002phq15,saunders1993audit}; the alliance; the coding of clinical interaction (MITI\cite{moyers2016miti} and VR-CoDES\cite{zimmermann2011vrcodes,delpiccolo2011vrcodes}); personality (IPIP\cite{goldberg1992markers,goldberg2006ipip} and TIPI\cite{gosling2003tipi}), which we have previously tracked in psychotherapy dialogue\cite{lin2025personality}; well-being\cite{topp2015who5}, self-esteem\cite{schmitt2005rses}, perceived stress\cite{cohen1983pss}, loneliness\cite{russell1996ucla,hughes2004ucla3,dejonggierveld2006djg}, decentering\cite{fresco2007eq}, rumination\cite{treynor2003rrs}, self-compassion\cite{raes2011scssf} and emotion regulation\cite{gross2003erq}; and the construct definitions of the Research Domain Criteria\cite{insel2010rdoc,cuthbert2013rdoc}.

With both embedders, items organized by construct domain (Fig.~\ref{fig:banks}a). For a median bank, the nearest neighbour of \nMbNNDomMedian{} of its items lay in another domain (MiniLM; \nMbNNDomMedianBge{} for bge). The exceptions were scales written as short adjectives or statements of well-being: the adjective pairs of TIPI (\nMbNNDomTIPI{} of items) mostly neighboured the one-word item names of clinician rating scales, and the items of WHO-5 (\nMbNNDomWHO{}) neighboured personality and psychosis-proneness items. Cross-instrument neighbours within the same domain were common, spanning parallel forms, revisions and short versions (every PHQ-8 item's nearest neighbour was its PHQ-9 counterpart) to distinct instruments for the same condition (GAD-7: \nMbNNBankGAD{} of items; Fig.~\ref{fig:banks}c). Keeping each of the \nMbDupTexts{} texts that occur in more than one bank only once left the median shares unchanged. The \nRdocDefs{} construct definitions of the Research Domain Criteria, written to cut across diagnoses, still clustered by their domain: their nearest neighbour lay in the same domain for \nRdocNNMini{} of definitions, against \nRdocChance{} expected by chance (Extended Data Fig.~\ref{fig:edrdoc}).

Banks occupied few effective dimensions (bge: \nMbBgeRawMin{}--\nMbBgeRawMax{}), largely because embeddings share a dominant direction\cite{ethayarajh2019contextual,mu2018allbutthetop}. Removing it raised dimensionality to at most \nMbBgeMRMax{} (Fig.~\ref{fig:banks}b). Instruments thus bring their own low-dimensional, domain-structured geometry to language scoring, where it reappears in the item scores. These clusters arise from descriptions of constructs before any participant response enters the analysis. Their interpretability makes them useful for comparing instruments and simultaneously makes them a plausible source of apparent psychological structure when the same texts are used to score speech.

\section*{Discussion}

Psychometric representational similarity analysis places sources of psychometric information in a common format. COMPASS 2.0 turns these representations into construct scores. Their first lesson is that an instrument contributes structure before a person contributes evidence. Wording dominated language-derived symptom geometry, and interview scripts imposed further structure. Beyond both, the registered test detected no population structure matching self-report, while one scorer's small exploratory residual depended on the reference. Speech conveyed general distress moderately well and specific symptoms much more faintly. Earlier keyed Bond scores likewise tracked relationship talk. Stance scoring distinguished counselling quality in a second corpus, but topic measures performed better in discovery and stance's advantage was not robust. Session topics, studied through psychotherapy topic models\cite{lin2023topic}, can themselves carry quality information.

The second lesson concerns validation. Population RDMs remain unchanged when self-reports are assigned to the wrong speakers. Their agreement shows that language organizes items as self-report does, not that it measures the right person. RDMs can also describe one speaker's language over time. Person-level validity requires paired measures, including COMPASS scores and item correspondence. Here, those measures tracked distress and vanished under shuffling. Validation needs both layers, with a clear account of the question each answers. The shuffle makes a property of the estimand visible: population geometry deliberately discards person identity to describe relationships among items. It is well suited to that question. Increasing its correlation with a target cannot supply a correspondence between people that the comparison never used. Validation must preserve the distinctions the intended interpretation depends on.

Language retains clinical information, but language-derived item structure cannot be accepted at face value. Symptom networks, factor structures and profiles, increasingly preferred to sum scores\cite{fried2015sumscores,fried2017fiftytwo}, require a wording baseline. Report geometric agreement with item texts, centre activities on eliciting questions, and test residual structure against a declared reference. Examine alternatives such as within-speaker covariance under their stated assumptions. Pair person-level read-outs with person shuffles, and interpret keyed cosine sums as topic measures unless the scorer distinguishes stance. This procedure needs item texts, scored language and, where available, item-level self-reports. It applies unchanged across instruments and taxonomies such as the Research Domain Criteria, across sessions and moving windows, and to two speakers' geometric alignment within conversations. Clinicians' ratings, theories and simulated respondents can supply further RDMs against which language representations and their COMPASS scores are checked.

Two literatures meet here. Survey research shows that item wording predicts response correlations\cite{arnulf2014predicting,rosenbusch2020semantic,wulff2025semantic,hommel2025language,pellert2026values}, suggesting that questionnaire measurement is partly semantic. This semantic structure is informative about how instruments organize their subject matter. Our question begins when those descriptions are used as probes to assess a new speaker. Language-based assessment then faces the converse hazard: text-derived scores inherit text geometry without person-level signal, and shared wording can produce agreement with self-report. Neuroscience's treatment of stimulus and noise structure provides an analogy\cite{walther2016reliability,diedrichsen2021wuc}. Probe whitening rescales item coordinates by reference covariance rather than weighting dissimilarity-estimate precision. Its validity therefore depends on the reference's account of nuisance. Within-person and between-person structure need not coincide\cite{molenaar2009new,adolf2014measurement}, so no reference is universally appropriate. Removing a pattern is not sufficient evidence that it was a confound. The assumptions that make a covariance a nuisance reference must also distinguish it from the signal one hopes to retain. The simulations make this requirement concrete: a seemingly stronger correction can erase genuine structure or reverse its sign.

Several limits matter. Samples were modest (\nDaicPersons{}, \nEdaicPersons{} participants), with one interview each and short answers (median \nDaicWordsPerWin{}, \nEdaicWordsPerWin{} words per window). Automatic confirmation transcripts add another source of error. Small item-specific effects could be missed, and individual geometry remained beyond reach. Registered-test power was high only for strong structure (\nFLPowerFourN{}\% at $g=0.4$, \nFLPowerTwoN{}\% at $g=0.2$ for 189 simulated people). Additive Gaussian simulations omit normalized embeddings, ordinal self-reports and estimated targets from the full pipeline. Whitening diagnoses departures from a reference rather than estimating personal structure, with inference conditional on the 25 fixed items. About a fifth of windows contained only one or two words, scored without their question. Weak scores therefore do not establish absent symptom information in the conversation. Larger models may read context and stance better than our compact local models, but their item-worded prompts still require the same null. Alliance analyses used demonstration-session quality, not alliance ratings. Four-turn exchanges and third-person statements may provide insufficient context, and contrasting corpus results warrant direct alliance ratings, longer exchanges and larger models. Sensitivity analyses only partly address repeated videos and shared material. Response styles, interviewer choices and setting-specific topics impose further structure.

The framework suggests a programme of measurement experiments. Longer language records paired with item-level questionnaires, such as histories of online posts, could support individual geometries. Direct alliance ratings would permit comparison of topic and stance against alliance itself. Further tools from representational similarity analysis include cross-validated distances for unbiased person-level geometries\cite{walther2016reliability}, resampling items as well as people to generalize beyond the instrument, and flexible models estimating how speech weights item features\cite{diedrichsen2017representational}. Topological comparisons emphasizing neighbourhoods over noisier long distances may improve robustness\cite{lin2024topology}. Acoustic and visual representations offer further applications, with identity and recording channel imposing structure as wording does here. A further test would vary probe wording while holding a person's language fixed, then vary the speaker while holding the probes fixed. Together, these interventions could reveal which features of an assessment travel with the instrument and which travel with the person.

A questionnaire is a theory of a person written in sentences. A model reading people through those sentences can return the theory as though it were the person. Psychometric representational similarity analysis places both in one space, allowing us to ask which is speaking. The promise of language-based assessment rests on this distinction: human expression may exceed the categories through which we choose to read it. A convincing measure must leave room for what the person says to change what the instrument leads us to expect.

\section*{Methods}

\paragraph{Data.} DAIC-WOZ\cite{gratch2014daic,devault2014simsensei} contains semi-structured clinical interviews conducted by a virtual human controlled by a human operator, with human transcripts and speaker turns. We used the \nDaicPersons{} interviews with transcripts (participants 300--492). E-DAIC\cite{ringeval2019avec} extends the corpus with interviews conducted by an autonomous agent; we used participants 600--718 (\nEdaicPersons{} interviews), for which automatic transcripts of the participant's speech are distributed. Self-reported PHQ-8 items\cite{kroenke2009phq8} (0--3) and PCL-C items\cite{blanchard1996pcl} (1--5; one response in the confirmation sample was coded 0) are available for all participants. Both corpora are available from the University of Southern California Institute for Creative Technologies under a data use agreement, and all processing was local. The psychotherapy sessions came from a licensed collection of transcribed counselling and psychotherapy sessions (Alexander Street), deduplicated by content to \nAlexSessions{} unique sessions with \nAlexTurns{} turns; sessions carry collection tags (anxiety \nCompAnx{}, depression \nCompDep{}, schizophrenia \nCompScz{}, suicidal \nCompSui{}; \nCompMulti{} sessions with several tags) but no outcome labels. AnnoMI\cite{wu2023annomi} contains \nAnnomiSessions{} transcribed motivational-interviewing demonstrations labelled as high or low quality. The high- and low-quality counselling corpus\cite{perezrosas2019counselor} contains \nHlqcSessions{} transcribed sessions with the same labels. Our loader set aside \nHlqcShared{} of them as shared with AnnoMI: \nHlqcSharedID{} share a source video identifier, and \nHlqcUnresolved{} had no resolvable identifier and were treated as shared by an error in the loader. This left \nHlqcIndep{} sessions from \nHlqcVideos{} videos for confirmation; a later transcript check found that \nHlqcTextDependent{} of them were AnnoMI material under another identifier (Supplementary Note 2). Before this study, the interview corpora had been used for other analyses of depression ratings; no item-level self-report was read by the code of this study before the registration, and no quality label was compared with any score before the alliance registration.

\paragraph{Items and probe banks.} The symptom items were the 8 PHQ-8 and 17 PCL-C items. The alliance items were the 36 items of the client, therapist and observer forms of the Working Alliance Inventory\cite{horvath1989wai,tichenor1989wai}, scored on Task, Bond and Goal with \nWaiReverse{} reverse-keyed items, and the client-form wording used by our earlier method\cite{lin2025compass}. The further banks were the MITI 4 behaviour codes\cite{moyers2016miti}, the VR-CoDES cue, concern and response definitions\cite{zimmermann2011vrcodes,delpiccolo2011vrcodes}, the 50-item IPIP Big-Five markers\cite{goldberg1992markers,goldberg2006ipip}, TIPI\cite{gosling2003tipi}, the UCLA Loneliness Scale version 3 and its three-item form\cite{russell1996ucla,hughes2004ucla3}, the six-item De Jong Gierveld scale\cite{dejonggierveld2006djg}, WHO-5\cite{topp2015who5}, the Rosenberg Self-Esteem Scale\cite{schmitt2005rses}, the Perceived Stress Scale\cite{cohen1983pss}, the decentering items of the Experiences Questionnaire\cite{fresco2007eq}, the brooding and reflection items of the Ruminative Responses Scale\cite{treynor2003rrs}, the Self-Compassion Scale short form\cite{raes2011scssf}, the Emotion Regulation Questionnaire\cite{gross2003erq}, clinical inventories (PHQ-9, GAD-7, PCL-5, the Columbia Suicide Severity Rating Scale, K10, PHQ-15, AUDIT, PQ-B and CAPE-42\cite{kroenke2001phq9,spitzer2006gad7,blevins2015pcl5,posner2011cssrs,kessler2002k10,kessler2003k6,kroenke2002phq15,saunders1993audit,loewy2011pqb,stefanis2002cape}, the BDI-II item titles\cite{beck1996bdi}, and the item definitions or names of HAM-A, BPRS, PANSS, SANS and SAPS\cite{hamilton1959hama,overall1962bprs,guy1976ecdeu,kay1987panss,andreasen1982sans,andreasen1989sans}) and the \nRdocDefs{} construct and subconstruct definitions of the Research Domain Criteria matrix in \nRdocDomains{} domains\cite{insel2010rdoc,cuthbert2013rdoc}. Texts were taken verbatim from the published sources (Supplementary Table~\ref{tab:banks}), instruction stems were excluded and each code definition was represented once; copyrighted texts were used for analysis only and are not redistributed.

\paragraph{Speech units.} In DAIC-WOZ, consecutive participant lines between two interviewer lines formed one answer window (\nDaicWindows{} windows; median \nDaicWinPerPerson{} per participant), grouped by the most recent interviewer question (\nDaicQuestions{} questions). In E-DAIC, recognized segments separated by pauses shorter than 3 s were merged into windows (\nEdaicWindows{}; median \nEdaicWinPerPerson{} per participant), grouped by decile of session time. In psychotherapy and counselling sessions, the unit was the speaker turn. Annotations in brackets were removed.

\paragraph{Scorers.} Cosine scorers embedded speech units and item texts with all-MiniLM-L6-v2\cite{reimers2019sbert,wang2020minilm} (revision 1110a243) or bge-large-en-v1.5\cite{xiao2024cpack} (revision d4aa6901), normalized the embeddings and took inner products. The entailment scorer was nli-deberta-v3-base\cite{he2023debertav3,williams2018mnli,bowman2015snli} (revision 6c749ce3). The premise was the answer window and the hypotheses were \nNliAtoms{} atomic first-person statements of the \nSymItems{} items (items naming several symptoms were split into parts); an item's score was the maximum over its parts of $P(\text{entailment})-P(\text{contradiction})$. On neutral sentences the scorer gave a mean of \nNliNeutral{}, and scoring the item texts against the atoms gave a mean of \nNliKdiag{} on the same item and \nNliKoff{} on different items; this probe-confusion matrix $K$ served as a second control for the entailment scorer. All models ran locally on CPU with pinned revisions.

\paragraph{Correspondence with representational similarity analysis.} Psychometric representational similarity analysis follows the logic of representational similarity analysis\cite{kriegeskorte2008rsa}. The items play the role of the experimental conditions, and each source supplies its own measurement channels: speech units or participants' mean profiles for language, participants for self-report, and embedding dimensions for item texts. A psychometric RDM is computed for each source over the same conditions, theories enter as model RDMs, and sources are compared by the second-order similarity of their RDMs. The covariance $C$ (equation~(\ref{eq:geometry})) is the centred second-moment matrix of the item activity patterns, from which distances follow: $C_{jj}+C_{kk}-2C_{jk}$ is the squared Euclidean distance between the centred activity patterns of items $j$ and $k$ (scaled by $1/(T-1)$), and $1-r_{jk}$ is half the squared distance between their standardized patterns. Trait items have served as conditions in neural representational similarity analysis\cite{izuma2025selfref}; here the representations come from language, self-report, item texts and theory. The wording null is a feature model in the sense of representational model analysis\cite{diedrichsen2017representational}: item activities are the item embeddings weighted by speech, so that $C=A^{\top}\Sigma A$, and an isotropic prior on the weights predicts $C\propto M$. Whitening expresses speech covariance in an orthonormal basis of the item embeddings, removing the feature model's prediction in a step analogous to multivariate noise normalization\cite{walther2016reliability}. Item permutation is the condition-label randomization test of representational similarity analysis\cite{kriegeskorte2008rsa,vandenbosch2025rsatoolbox}. RDMs were compared by Spearman correlation over item pairs; Kendall's $\tau_A$, which is recommended when a model RDM has many ties, ordered the comparisons in Fig.~\ref{fig:hub}b in the same way (rank agreement \nHubTauAgree{}). The split-half reliability of each sampled RDM across random halves of participants, $r_1$, bounds its attainable correlation with another source of reliability $r_2$ at about $\sqrt{r_1r_2}$ under classical assumptions, in the spirit of the noise ceiling of representational similarity analysis\cite{vandenbosch2025rsatoolbox}.

\paragraph{Wording null and whitening.} Let $A$ hold the unit-norm item embeddings as columns and $M=A^{\top}A$ be the item-text similarity. A speech unit with embedding $\mathbf{e}$ has item activities $\mathbf{p}=A^{\top}\mathbf{e}$, so $\mathrm{Cov}(\mathbf{p})=A^{\top}\Sigma A$, which equals $\sigma^2 M$ when $\Sigma=\sigma^2I$. If people's standardized item levels $\mathbf{z}$ ($\mathrm{Cov}(\mathbf{z})=R$) are expressed in speech as $\mathbf{e}=gA\mathbf{z}+\mathbf{u}$, with $\mathrm{Cov}(\mathbf{u})=\sigma^2I$ and $\mathrm{Cov}(\mathbf{z},\mathbf{u})=0$, then $\mathrm{Cov}(\mathbf{p})=g^2MRM+\sigma^2M$, so the observed geometry mixes the person's structure with the wording, and the wording dominates when $g$ is small. The whitened geometry (equation~(\ref{eq:whiten})), with $C$ the covariance of person profiles, has zero off-diagonal entries under the null and, under the model, equals $g^2M^{1/2}RM^{1/2}+(\sigma^2-1)I$, so that its off-diagonal entries are those of $g^2M^{1/2}RM^{1/2}$. We compared $W$ with self-reported correlations by partial correlation over item pairs, controlling for $M$. Exact deconvolution, $M^{-1}CM^{-1}=g^2R+\sigma^2M^{-1}$ with $g^2$ and $\sigma^2$ fitted on the diagonal ($R_{jj}=1$), recovers $R$ in the population when $g>0$ and the diagonal regression is identifiable; in samples its fitted variances can be non-positive, so we report the share of inadmissible fits, the performance of admissible fits and that of a positive-semidefinite projection. Ridge regularization requires a different correction and was not used. The isotropic reference requires equal speech variance in every direction. As an empirical alternative we whitened by the pooled within-speaker covariance of the centred window activities, $S_w=\sum_i\sum_t(\mathbf{p}_{it}-\bar{\mathbf{p}}_i)(\mathbf{p}_{it}-\bar{\mathbf{p}}_i)^{\top}/\sum_i(n_i-1)$, which captures topic and other structure in each speaker's own speech and applies to any scorer, including the entailment model, for which $M$ (the mean of the two embedders' similarity matrices) is a heuristic rather than a derived reference. Whitening by $S_w$ tests for person structure only if between-speaker nuisance covariance is proportional to $S_w$ and within-speaker variation carries no symptom structure; the simulations show that it fails otherwise, so we treat it as a sensitivity analysis. We compared it with the registered analysis three ways: controlling for Corr$(S_w)$ only, for the registered controls only (so that only the reference changes) and for both.

\paragraph{Profiles and centring.} Item activities of each window were centred on the mean of the windows answering the same question (or in the same phase) from all other participants, so that no participant's own windows entered their centring, except for \nSingletonWindows{} windows in questions answered by a single participant, which were centred on their own mean. A participant's profile was the mean of their centred windows; naive profiles used the uncentred activities, as in standard practice. Language geometry was the correlation (naive) or covariance (for whitening) of profiles across participants; self-reported geometry was the correlation of item scores across the same participants. For the entailment scorer, $M$ was the mean of the two embedders' similarity matrices.

\paragraph{Registered tests.} H1: $r(\text{naive geometry},M)-r(\text{naive geometry},\text{self-report})>0$ for the cosine scorers, supported if the person-bootstrap interval (2,000 resamples) lay above zero; 95\% in discovery and 97.5\% in confirmation (Bonferroni over two scorers). H2 (primary): the partial correlation over item pairs between $W$ and the self-reported correlations, controlling for $M$ (and $K$ for the entailment scorer), was positive with a Freedman--Lane item-permutation $P<0.05$ (10,000 permutations\cite{freedman1983nonstochastic,winkler2014permutation}), Holm-adjusted across scorers in confirmation. H3: across participants, the mean correlation of each language-derived item score with the same self-reported item minus its mean correlation with the other items (both z-scored) was positive, by person bootstrap; 95\% in discovery and 98.3\% in confirmation (Bonferroni over three scorers). Discovery was analysed first; confirmation was analysed once with unchanged code. Item permutation, rather than permutation of matrix entries, respects the dependence among item pairs\cite{mantel1967detection,guillot2013dismantling}.

\paragraph{Simulation.} Twenty-five unit-norm item vectors in 128 dimensions formed eight wording clusters, three of which held one item from each of two instruments. True item correlations mixed a wording-aligned part (items loading on their cluster) and a wording-free part (three random factors) in equal measure, which gave a correlation of \nSimTruthWording{} between the true and wording geometries. Each simulated person had item levels drawn with these correlations, a speech mean $gA\mathbf{z}$ plus person-level variation unrelated to the items, and 155 windows of independent noise; activities were projected onto the items. We ran 200 samples of 189 people for $g\in\{0,0.1,0.2,0.4\}$ and recorded the correlations of the naive, whitened and exactly deconvolved geometries with the truth and with the wording. A further simulation compared the Freedman--Lane test after whitening by $M$ or by the within-speaker covariance (55 windows per speaker; 200 samples without and 100 with person signal, so that a false-positive rate of 5\% has a Monte Carlo standard error of about 1.5 percentage points; 499 permutations) in four conditions: topic variation of the same shape between and within speakers; stronger between-speaker topic variation of the same shape; between- and within-speaker topic variation along different item-cluster directions; and genuine within-speaker item dynamics with the same correlation structure as the person levels, without topic variation. These simulations are stylized additive Gaussian models, not reproductions of the full pipeline. The false-positive rate and power of the registered Freedman--Lane test were estimated in separate runs (\nFLReps{} samples at $g=0$ and 100 at $g=0.2$ and $0.4$, for 189 and 86 people; 999 permutations each). A second simulation added question-specific content and different question subsets per person, and a third generated forward and reverse alliance items sharing topic wording.

\paragraph{General factor (exploratory).} For each sample and scorer, language-derived profiles and self-reported items were z-scored per item. The general factors were the first principal components of each, signed to correlate positively with the item mean. Item-specific convergence was computed as in H3 after projecting the self-reported general factor out of both sets of item scores. Intervals came from 2,000 person-bootstrap resamples.

\paragraph{Alliance re-examination.} For each psychotherapy turn, the keyed subscale score was the sum over the subscale's items of the item's key ($+1$ or $-1$) times the cosine between the turn and the item (client form for client turns, therapist form for therapist turns, as in our earlier method), and the topic activation was the mean cosine with the subscale's items regardless of key. Session scores were means over turns. We report the variance in keyed scores explained by topic activation across sessions, the association of therapist-minus-client differences, and the variance explained by the clinical collection among sessions with a single tag.

\paragraph{Pre-specified alliance tests.} Each session was cut into consecutive four-turn segments, written as a dialogue and scored against the 36 observer-form items by the entailment model as $P(\text{entailment})-P(\text{contradiction})$. The stance score of a session was the mean over segments of the keyed item mean. The keyed cosine score applied our earlier rule to the same items: the mean over turns of the keyed mean cosine (MiniLM). The primary tests, in a fixed sequence at $\alpha=0.05$, were: (A1) the area under the receiver operating characteristic curve (AUC) of the stance score for high- versus low-quality sessions exceeds 0.5, by a stratified-bootstrap 95\% interval (5,000 resamples within quality groups); (A2) if A1 holds, the stance AUC exceeds the keyed cosine AUC, by a paired stratified-bootstrap interval. Confirmation decided; discovery was reported with the same statistics. Subscale AUCs with label-permutation tests (Holm across subscales), a replica of our earlier client and therapist scoring, and correlations with topic activation and session length were secondary.

\paragraph{Probe-bank geometry.} For each bank and embedder we computed the participation ratio $(\sum_i\lambda_i)^2/\sum_i\lambda_i^2$ of the eigenvalues of the item-similarity matrix\cite{gao2017dimensionality}, before and after subtracting the mean embedding of all items, the share of items whose nearest neighbour (by cosine) lay in another bank or another construct domain, and a classical multidimensional scaling of the cosine distances $\sqrt{2-2\cos}$.

\paragraph{Sensitivity analyses (exploratory).} The following checks were conducted after the registered analyses. Person shuffle: self-reports were reassigned among participants 1,000 times and the geometric and person-paired statistics recomputed. Effect bounds: person-bootstrap intervals (2,000 resamples) for the H2 partial correlations, computed conditionally on the estimated question means. Exchangeability: Freedman--Lane tests with item permutations restricted within the PHQ-8 and the PCL-C, using the registered controls ($M$, and $K$ for the entailment scorer; 10,000 permutations); the observed statistic equals the registered one. Within-speaker reference: H2-type tests after whitening by $S_w$ under the three control sets (5,000 permutations; 1,000 bootstrap resamples with $S_w$ re-estimated in each, conditional on the estimated question means). Out-of-range response: the confirmation tests without participant \nPCLZeroPerson{}. Baselines for the person-paired read-outs: partial correlations after removing the ranks of the number of answer windows, the number of participant words and topic activation (the mean MiniLM cosine similarity of a participant's windows to the 25 symptom items), none of which is fitted to the self-reports (2,000 person-bootstrap resamples). Alliance: bootstrap over whole source videos (5,000 resamples); a transcript screen of every session against every AnnoMI session (the share of the session's word 5-grams contained in one AnnoMI session; text match at \nTextThreshold{} or more, one threshold throughout; within the registered confirmation sample the observed values were at most \nGapBelow{} or at least \nGapAbove{}, a gap that does not hold for all excluded sessions), with every session classified by source-identifier status and screen (Supplementary Table~\ref{tab:provenance}); and stance scores recomputed without segments in which any premise--hypothesis pair exceeded the 256-token limit (\nAlTruncShare{}\% of the \nAlTruncSegments{} segments of the registered confirmation sample).

\paragraph{Ethics and data use.} This study analysed existing interview and psychotherapy transcripts and previously collected assessments. It involved no new recruitment or interaction with participants. Access to DAIC-WOZ and E-DAIC is governed by a data use agreement, and the Alexander Street psychotherapy transcripts require licensed access. The reviewer code package contains aggregate results and does not redistribute these transcripts.

\paragraph{Pre-registration.} The symptom tests were pre-registered on OSF Registries (\url{https://osf.io/j9q26/}) before any self-reported item score was read, and the scoring and analysis code were fixed at the same time. The alliance tests were specified in a second analysis plan, with their scoring and analysis code fixed, before any session-quality label was compared with model output. Deviations are listed in Supplementary Note 2.

\paragraph{Statistics.} Intervals are percentile bootstrap intervals over the resampling units (participants, sessions or, in sensitivity analyses, source videos); item-level inference treats the 25 items as fixed. Correlations are Pearson correlations unless stated otherwise. Split-half reliabilities were corrected to full length with the Spearman--Brown formula\cite{spearman1910correlation,brown1910correlation}. Effective dimensions, reliabilities and correlations with wording are descriptive. The random seed of every registered analysis was 20261003.

\section*{Data availability}
DAIC-WOZ and E-DAIC are available from the University of Southern California Institute for Creative Technologies under a data use agreement. The psychotherapy transcripts are available by subscription from Alexander Street (Counseling and Psychotherapy Transcripts). AnnoMI and the high- and low-quality counselling corpus are publicly available from their authors\cite{wu2023annomi,perezrosas2019counselor}. Item texts in the public domain are distributed with the code; other item texts are available from their publishers. The code package distributes aggregate results only.

\section*{Code availability}
Code for reproducing every number, table and figure, including the COMPASS 2.0 implementation of psychometric representational similarity analysis, is provided in xpsych, a Python package for explainable psychiatry (\url{https://github.com/linlab/xpsych}). The dedicated \texttt{papers/compass2} folder contains the code and instructions for reproducing this study. Reproduction requires access to the source datasets under their respective access conditions.

\section*{Acknowledgements}
This work was supported in part by a 2025 NARSAD Young Investigator Grant from the Brain \& Behavior Research Foundation, funded by The Sidney R. Baer, Jr. Foundation (Grant ID 33941). B.L. is also supported by the U.S. National Institutes of Health (NIH), the U.S. Department of Veterans Affairs (VA), the Hasso Plattner Foundation, and the Windreich Family Foundation. We thank the University of Southern California Institute for Creative Technologies for making the interview corpora available.

\section*{Author contributions}
B.L. conceived the study, performed the analyses, and wrote the manuscript.

\section*{Competing interests}
The author declares no competing interests.

\clearpage
\section*{Extended Data}
\renewcommand{\figurename}{Extended Data Fig.}
\renewcommand{\tablename}{Extended Data Table}
\renewcommand{\theHfigure}{ED\arabic{figure}}
\renewcommand{\theHtable}{ED\arabic{table}}
\setcounter{figure}{0}
\setcounter{table}{0}

\begin{figure}[h]
  \centering
  \includegraphics[width=\textwidth]{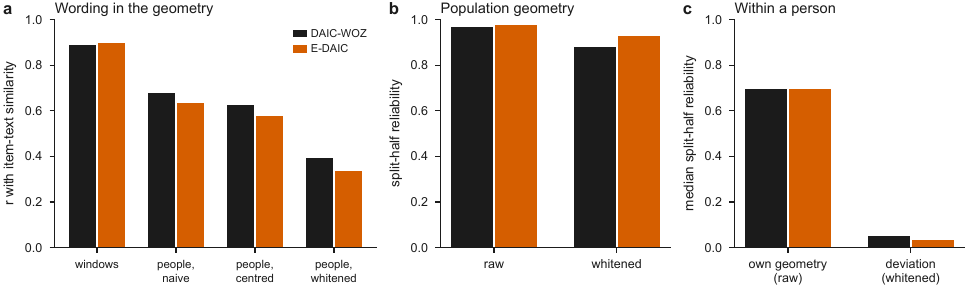}
  \caption{\textbf{Script structure and reliability in both interview corpora.} \textbf{a}, Correlation of the language-derived item geometry with item-text similarity at the level of single windows and of participants (naive, question- or phase-centred, and centred and whitened); MiniLM. \textbf{b}, Split-half reliability (Spearman--Brown) of the population geometry, raw and whitened. \textbf{c}, Median split-half reliability, across random halves of each participant's windows, of the participant's own window-level item geometry (raw correlations, which share the wording) and of that geometry whitened against the other participants' window covariance (the participant's deviation); participants with at least 40 windows.}
  \label{fig:edreliab}
\end{figure}

\begin{figure}[h]
  \centering
  \includegraphics[width=\textwidth]{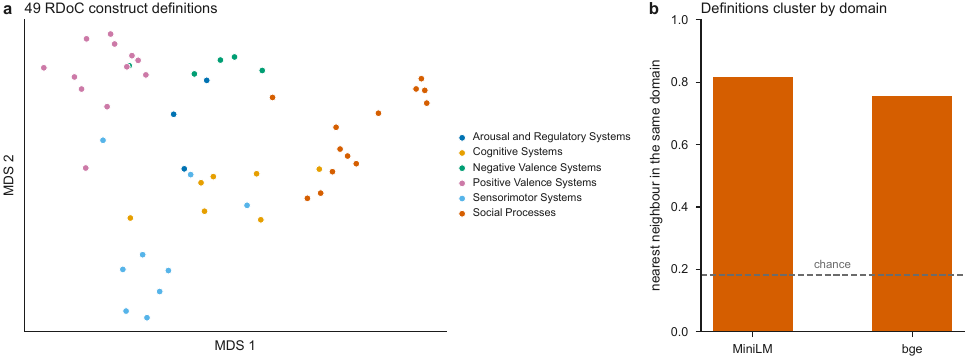}
  \caption{\textbf{Construct definitions of the Research Domain Criteria cluster by domain in embedding space.} \textbf{a}, Classical multidimensional scaling of the cosine distances between the \nRdocDefs{} construct and subconstruct definitions (MiniLM), coloured by domain. \textbf{b}, Share of definitions whose nearest neighbour lies in the same domain for each embedder, against the share expected by chance.}
  \label{fig:edrdoc}
\end{figure}

\begin{figure}[h]
  \centering
  \includegraphics[width=\textwidth]{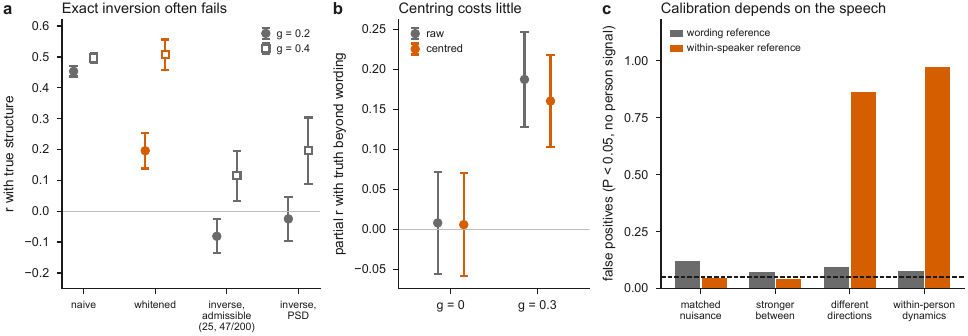}
  \caption{\textbf{Simulation details.} \textbf{a}, Correlation with the true symptom structure of the naive geometry, the whitened geometry, exact deconvolution restricted to admissible fits (positive fitted signal variance and positive noise-subtracted diagonal; their number out of 200 is given) and a positive-semidefinite projection of all deconvolution fits, at $g=0.2$ (circles) and $g=0.4$ (squares); mean $\pm$ s.d.\ over 200 samples of 189 people. Whitening does not estimate the true structure itself, so this panel compares structural agreement. \textbf{b}, Partial correlation with the truth beyond wording of person-level geometry when people answer different subsets of questions, without and with centring on the question mean, at $g=0$ and $g=0.3$: centring keeps the null calibrated and costs little power. \textbf{c}, False-positive rate of the Freedman--Lane test without person signal after whitening by the wording reference $M$ or by the within-speaker covariance, in four conditions: topic variation of the same shape between and within speakers, stronger between-speaker topic variation of the same shape, topic variation along different directions between and within speakers, and genuine within-speaker item dynamics (200 samples each; dashed line, 5\%).}
  \label{fig:edsim}
\end{figure}

\begin{table}[h]
\centering\small
\caption{\textbf{Corpora.} Units are answer windows (interviews) or speaker turns (sessions).}
\label{tab:corpora}
\begin{tabularx}{\textwidth}{@{}lXrrl@{}}
\toprule
Corpus & Content & People or sessions & Units & Use \\
\midrule
DAIC-WOZ & Clinical interviews by a human-operated virtual interviewer; human transcripts & 189 & 10,533 & Discovery (symptoms) \\
E-DAIC & Interviews by an autonomous agent; automatic transcripts & 86 & 5,214 & Confirmation (symptoms) \\
Alexander Street & Transcribed psychotherapy sessions, four clinical collections & 786 & 180,547 & Alliance re-examination \\
AnnoMI & Motivational-interviewing demonstrations, high/low quality (110/23) & 133 & 9,687 & Discovery (alliance) \\
High/low-quality counselling & Counselling sessions not in AnnoMI, high/low quality (119/65) & 184 & 7,877 & Confirmation (alliance) \\
\bottomrule
\end{tabularx}
\end{table}

\begin{table}[h]
\centering\small
\caption{\textbf{Registered symptom tests.} H1: correlation of the naive language geometry with item-text similarity / with self-reported item correlations (interval of the difference: 95\% in discovery, 97.5\% in confirmation). H2: partial correlation of the whitened geometry with self-report beyond wording (Freedman--Lane $P$; Holm-adjusted in confirmation). H3: mean same-item minus different-item correlation (95\% or 98.3\% interval).}
\label{tab:registered}
\begin{tabularx}{\textwidth}{@{}llXll@{}}
\toprule
Sample & Scorer & H1 & H2 & H3 \\
\midrule
Discovery & MiniLM & 0.68 / 0.42 (0.15 to 0.47) & \ensuremath{-}0.009 ($P$ = 0.93) & 0.004 (\ensuremath{-}0.016 to 0.023) \\
 & bge & 0.69 / 0.47 (0.15 to 0.40) & 0.028 ($P$ = 0.52) & 0.014 (\ensuremath{-}0.002 to 0.031) \\
 & Entailment & -- & 0.051 ($P$ = 0.27) & 0.038 (0.023 to 0.056) \\
Confirmation & MiniLM & 0.63 / 0.36 (0.11 to 0.55) & 0.008 ($P$ = 0.93) & 0.038 (0.007 to 0.072) \\
 & bge & 0.65 / 0.21 (0.33 to 0.58) & \ensuremath{-}0.096 ($P$ = 0.15) & 0.042 (0.013 to 0.073) \\
 & Entailment & -- & 0.118 ($P$ = 0.15) & 0.005 (\ensuremath{-}0.022 to 0.035) \\
\bottomrule
\end{tabularx}
\end{table}

\clearpage
\section*{Supplementary Information}
{\large\PaperTitle\par}
Baihan Lin
\medskip
\renewcommand{\figurename}{Supplementary Fig.}
\renewcommand{\tablename}{Supplementary Table}
\renewcommand{\theHfigure}{S\arabic{figure}}
\renewcommand{\theHtable}{S\arabic{table}}
\setcounter{figure}{0}
\setcounter{table}{0}

\subsection*{Supplementary Note 1: the wording null and its whitening}
\label{sn:null}

Let $A\in\mathbb{R}^{D\times k}$ hold the unit-norm embeddings of $k$ item texts and let a stretch of speech have embedding $\mathbf{e}\in\mathbb{R}^{D}$. Cosine scoring gives the item activities $\mathbf{p}=A^{\top}\mathbf{e}$ (for normalized $\mathbf{e}$), so for any distribution of speech with covariance $\Sigma$,
\begin{equation*}
\mathrm{Cov}(\mathbf{p})=A^{\top}\Sigma A .
\end{equation*}
The probes therefore transform speech covariance by congruence. Two limiting cases show what this means for measurement. If speech is isotropic in the embedding space, $\Sigma=\sigma^2 I$ and $\mathrm{Cov}(\mathbf{p})=\sigma^2 M$ with $M=A^{\top}A$: the score correlations equal the cosine similarities of the item texts, whatever the people are like. If speech carries the person's item levels $\mathbf{z}$ (correlation $R$) through the item directions, $\mathbf{e}=gA\mathbf{z}+\mathbf{u}$ with $\mathbf{u}$ isotropic, then
\begin{equation*}
\mathrm{Cov}(\mathbf{p})=g^{2}MRM+\sigma^{2}M .
\end{equation*}
The person's structure thus reaches the scores through $M$ on both sides. Whitening by the wording,
\begin{equation*}
W=M^{-1/2}\,\mathrm{Cov}(\mathbf{p})\,M^{-1/2}-\sigma^{2}I=g^{2}M^{1/2}RM^{1/2},
\end{equation*}
removes the isotropic wording term exactly. Because $\sigma^2$ is unknown in practice, the test uses only the off-diagonal entries of $M^{-1/2}CM^{-1/2}$, which are zero under the null whatever $\sigma^2$ is. Under the model, the whitened geometry retains a residual congruence by $M^{1/2}$: it is a transformed image of $R$, not $R$ itself. This is why whitening serves as a diagnostic and the registered comparison with self-report also partials out $M$. With non-isotropic speech, the transform leaves structure whose origin need not be psychological. An empirical reference estimated from the speakers' own speech removes such structure only when between-speaker nuisance has the shape of within-speaker variation and within-speaker variation carries no symptom structure; otherwise it can create false positives or remove real structure (Methods; Extended Data Fig.~3c). Full deconvolution, $M^{-1}CM^{-1}=g^2R+\sigma^2M^{-1}$, recovers $R$ in the population once $g^2>0$ and $\sigma^2$ are fitted on the diagonal, but amplifies estimation error along the directions in which $M$ is small (its condition number is \nCondM{} for MiniLM). In samples of 189 people with $g=0.4$, \nInvFailDiag{} of 200 fits had a non-positive noise-subtracted variance and \nInvFailG{} a non-positive fitted signal variance, so the estimator is often undefined at the present sample size (Extended Data Fig.~3a).

The entailment scorer is nonlinear, but an empirical probe geometry permits a related diagnostic: scoring each item text against every hypothesis gives a probe-confusion matrix $K$ (mean \nNliKdiag{} on the diagonal and \nNliKoff{} off it), which the registered test used, together with $M$, as a control. $K$ is a probe-response matrix rather than a derived null covariance, so for this scorer whitening by $M$ is a heuristic. The empirical within-speaker reference (Methods) does not rely on $M$ or $K$, but it rests on assumptions of its own.

Interview scripts introduce another source of structure. If windows answer questions $q$ with question-specific content $\mathbf{c}_q$, then a person's mean activity includes $A^{\top}\bar{\mathbf{c}}$ over the questions they answered, so people who answered different questions differ in their profiles even without any symptom signal. Centring each window on the mean activity of its question over the other participants removes the estimated $A^{\top}\mathbf{c}_q$ without using the participant's own data (except where a question was answered by one participant only); question-by-person interactions remain.

\subsection*{Supplementary Note 2: deviations, corrections and sensitivity analyses}
\label{sn:deviations}

\paragraph{Registered analyses.} The symptom tests were run with the registered code unchanged. Because the entailment scores took longer to compute than the embeddings, the discovery analysis was first run for the cosine scorers alone, by restricting the registered code's list of scorers, and then for all three scorers; the cosine-scorer results of the two runs were identical. The confirmation analysis was run once, with all three scorers. The alliance tests were specified in a second analysis plan after the symptom tests had been analysed and before any session-quality label was compared with model output; the registered code for the stance scores was unchanged except for a setting of the number of processor threads, which does not affect the scores.

\paragraph{Corrections found after the registered analyses.} (i) The loader of the high- and low-quality counselling corpus mapped every missing or non-YouTube source address to the same empty identifier, which also occurred among the AnnoMI sources, so \nHlqcUnresolved{} sessions with a missing or non-YouTube source address (12 low and 9 high quality, one with a Vimeo address) were counted as shared with AnnoMI and excluded; \nHlqcSharedID{} sessions share a video identifier with AnnoMI. The registered confirmation sample (\nHlqcIndep{} sessions from \nHlqcVideos{} videos) is unchanged. A transcript screen then classified every session by source-identifier status and text match (Supplementary Table~\ref{tab:provenance}): \nHlqcTextDependent{} sessions of the registered confirmation sample matched AnnoMI material uploaded under another identifier, and \nHlqcUnresolvedNeg{} of the \nHlqcUnresolved{} excluded sessions were screen-negative. A negative screen does not establish independence, since sessions could share speakers, scripts or edited footage below the screen's resolution. (ii) One PCL-C response (participant \nPCLZeroPerson{}, hyperalertness item) was coded 0, outside the scale's range of 1--5; we did not recode it, and excluding the participant left every registered conclusion unchanged (entailment-scorer H2: $P=\nNoSixEightThreeNliHtwoP{}$ before multiplicity correction). (iii) In the probe-bank analysis, two AUDIT instruction stems and four duplicate VR-CoDES definitions had passed the filters; they were removed, and the reported bank results use the corrected set. (iv) Earlier simulations compared whitening with ridge-regularized deconvolution whose diagonal fit was valid only without the ridge; the comparison reported here uses exact deconvolution and reports how often its fits are inadmissible.

\paragraph{Sensitivity analyses.} These analyses followed the registered tests and are exploratory. Their $P$ values are not adjusted for multiplicity (Supplementary Tables~\ref{tab:sensitivity}--\ref{tab:provenance}). For symptoms, we computed person-bootstrap intervals for H2 and Freedman--Lane tests with within-instrument permutations and registered controls. We examined within-speaker whitening under three control sets, person shuffles (main text, Fig.~4f), and length and topic baselines for paired read-outs (Supplementary Table~\ref{tab:baselines}). For alliance, we bootstrapped source videos, excluded transcript matches to AnnoMI, added previously excluded screen-negative sessions and excluded truncated segments. Stance scores separated the quality groups in every set, whereas the lower bound of the interval for their advantage over cosine scores lay on either side of zero depending on the set.

\begin{table}[p]
\centering\scriptsize
\caption{\textbf{Sensitivity analyses of the symptom tests} (exploratory; $P$ values unadjusted). H2 interval: person-bootstrap 95\% interval of the registered partial correlation, conditional on the estimated question means. Block $P$: Freedman--Lane test of the registered statistic (controls $M$, and $K$ for the entailment scorer) with items permuted within the PHQ-8 and within the PCL-C. Within-speaker reference: the population geometry whitened by the pooled within-speaker covariance $S_w$ instead of $M$; $r(S_w,M)$ is the correlation of the reference geometry with item-text similarity; partial $r$ with self-report ($P$) controlling for Corr$(S_w)$ only, for the registered controls only (the change of reference alone), or for both. This is a sensitivity to the reference, valid as a test of person structure only if within-speaker variation is nuisance with the same shape as between-speaker nuisance (Extended Data Fig.~3c). D, discovery (DAIC-WOZ); C, confirmation (E-DAIC).}
\label{tab:sensitivity}
\begin{tabular}{@{}llllllll@{}}
\toprule
 & & & & & \multicolumn{3}{c}{Partial $r$ after whitening by $S_w$ ($P$), controlling for} \\
\cmidrule(l){6-8}
Sample & Scorer & H2 interval & Block $P$ & $r(S_w,M)$ & Corr$(S_w)$ & registered & both \\
\midrule
D & MiniLM & [$-$0.137, 0.129] & 0.925 & 0.904 & $-$0.027 (0.722) & 0.003 (0.972) & 0.017 (0.823) \\
D & bge & [$-$0.067, 0.109] & 0.520 & 0.877 & $-$0.031 (0.522) & 0.060 (0.190) & 0.059 (0.282) \\
D & entailment & [$-$0.012, 0.107] & 0.260 & 0.216 & 0.030 (0.567) & $-$0.005 (0.923) & $-$0.030 (0.551) \\
C & MiniLM & [$-$0.098, 0.123] & 0.962 & 0.914 & 0.029 (0.697) & 0.046 (0.522) & 0.076 (0.291) \\
C & bge & [$-$0.134, $-$0.001] & 0.057 & 0.889 & $-$0.079 (0.135) & $-$0.045 (0.367) & $-$0.004 (0.950) \\
C & entailment & [0.020, 0.162] & 0.031 & 0.193 & $-$0.011 (0.865) & $-$0.001 (0.993) & $-$0.047 (0.483) \\
\bottomrule
\end{tabular}
\end{table}

\begin{table}[p]
\centering\scriptsize
\caption{\textbf{Sensitivity analyses of the alliance tests} (exploratory). Intervals resample whole source videos (5,000 resamples). AUC, area under the curve for high- versus low-quality sessions; difference, stance minus keyed cosine. Screen-negative: no AnnoMI session contains at least 0.3 of the session's word 5-grams; this does not establish independence of actors, therapists or productions.}
\label{tab:alliancesens}
\begin{tabular}{@{}lrrlll@{}}
\toprule
Analysis & Sessions & Videos & Stance AUC [95\% CI] & Keyed cosine AUC & Difference [95\% CI] \\
\midrule
Discovery (AnnoMI) & 133 & 119 & 0.547 [0.420, 0.671] & 0.662 & [$-$0.275, 0.040] \\
Registered confirmation sample & 184 & 156 & 0.682 [0.583, 0.773] & 0.564 & [$-$0.009, 0.243] \\
Registered sample, screen-negative only & 176 & 150 & 0.695 [0.594, 0.785] & 0.550 & [0.013, 0.275] \\
All screen-negative sessions without ID match & 192 & 166 & 0.678 [0.585, 0.765] & 0.554 & [$-$0.0001, 0.246] \\
Registered sample without over-limit segments & 184 & 156 & 0.676 [0.581, 0.767] & 0.564 & [$-$0.019, 0.239] \\
\bottomrule
\end{tabular}
\end{table}

\begin{table}[p]
\centering\scriptsize
\caption{\textbf{Provenance of the high- and low-quality counselling sessions.} Left: all 258 sessions by registered status, source-identifier status and transcript screen (text match: at least 0.3 of the session's word 5-grams occur in one AnnoMI session). Right: the 21 sessions that the registered loader excluded without an identifier match (missing addresses, and one Vimeo address that the loader did not parse). In the registered sample, containment was at most 0.027 or at least 0.540.}
\label{tab:provenance}
\begin{minipage}[t]{0.46\textwidth}\centering
\begin{tabular}[t]{@{}lllr@{}}
\toprule
Registered & Identifier & Transcript & $n$ \\
\midrule
analysed & ID distinct & screen-negative & 176 \\
analysed & ID distinct & text match & 8 \\
excluded & ID distinct & screen-negative & 1 \\
excluded & ID match & screen-negative & 4 \\
excluded & ID match & text match & 49 \\
excluded & ID unresolved & screen-negative & 15 \\
excluded & ID unresolved & text match & 5 \\
\bottomrule
\end{tabular}
\end{minipage}\hfill
\begin{minipage}[t]{0.5\textwidth}\centering
\begin{tabular}[t]{@{}lllll@{}}
\toprule
Session & Source & Quality & Containment & Transcript \\
\midrule
high\_018 & none & high & 0.005 & screen-negative \\
high\_054 & none & high & 0.002 & screen-negative \\
high\_062 & none & high & 0.003 & screen-negative \\
high\_066 & none & high & 0.002 & screen-negative \\
high\_073 & none & high & 0.002 & screen-negative \\
high\_074 & none & high & 0.017 & screen-negative \\
high\_075 & none & high & 0.003 & screen-negative \\
high\_102 & none & high & 0.020 & screen-negative \\
high\_104 & vimeo:11671267 & high & 0.002 & screen-negative \\
low\_093 & none & low & 0.471 & text match \\
low\_094 & none & low & 0.705 & text match \\
low\_095 & none & low & 0.002 & screen-negative \\
low\_096 & none & low & 0.008 & screen-negative \\
low\_097 & none & low & 0.826 & text match \\
low\_098 & none & low & 0.683 & text match \\
low\_099 & none & low & 0.884 & text match \\
low\_100 & none & low & 0.003 & screen-negative \\
low\_101 & none & low & 0.003 & screen-negative \\
low\_102 & none & low & 0.003 & screen-negative \\
low\_103 & none & low & 0.004 & screen-negative \\
low\_104 & none & low & 0.002 & screen-negative \\
\bottomrule
\end{tabular}
\end{minipage}
\end{table}

\begin{table}[p]
\centering\scriptsize
\caption{\textbf{Person-paired read-outs beyond interview length and topic} (exploratory). COMPASS PHQ-8 score: mean of the participant's z-scored, question-centred PHQ-8 item scores, correlated with the self-reported PHQ-8 total; general factors: first principal components of the language-derived and self-reported item scores. Partial: after removing the ranks of the number of answer windows, the number of participant words and topic activation (mean MiniLM cosine similarity of the participant's windows to the 25 symptom items), none of which is fitted to the self-reports; log counts instead of ranks gave similar values. Lower panel: rank correlations of these baselines with the PHQ-8 total. Person-bootstrap 95\% intervals (2,000 resamples). D, discovery (DAIC-WOZ, 189 participants); C, confirmation (E-DAIC, 86).}
\label{tab:baselines}
\begin{tabular}{@{}llllll@{}}
\toprule
 & & \multicolumn{2}{c}{COMPASS PHQ-8 score with PHQ-8 total} & \multicolumn{2}{c}{General factors} \\
\cmidrule(lr){3-4}\cmidrule(l){5-6}
Sample & Scorer & $r$ & partial $r$ & $r$ & partial $r$ \\
\midrule
D & MiniLM & 0.15 [0.00, 0.30] & 0.07 [$-$0.08, 0.21] & 0.07 [$-$0.06, 0.22] & $-$0.04 [$-$0.16, 0.11] \\
D & bge & 0.26 [0.13, 0.37] & 0.16 [0.03, 0.29] & 0.26 [0.13, 0.38] & 0.14 [$-$0.01, 0.27] \\
D & entailment & 0.54 [0.44, 0.62] & 0.52 [0.43, 0.61] & 0.49 [0.38, 0.59] & 0.46 [0.35, 0.56] \\
C & MiniLM & 0.39 [0.24, 0.53] & 0.12 [$-$0.06, 0.28] & 0.33 [0.14, 0.50] & 0.02 [$-$0.26, 0.24] \\
C & bge & 0.53 [0.41, 0.65] & 0.40 [0.23, 0.56] & 0.52 [0.38, 0.65] & 0.36 [0.16, 0.52] \\
C & entailment & 0.57 [0.42, 0.68] & 0.49 [0.33, 0.63] & 0.60 [0.44, 0.72] & 0.54 [0.37, 0.66] \\
\midrule
Baseline & & windows & words & topic activation & \\
\midrule
D & & 0.19 [0.05, 0.33] & $-$0.04 [$-$0.18, 0.10] & 0.13 [$-$0.02, 0.27] & \\
C & & 0.21 [$-$0.02, 0.41] & 0.03 [$-$0.18, 0.25] & 0.34 [0.15, 0.52] & \\
\bottomrule
\end{tabular}
\end{table}

\subsection*{Supplementary Note 3: analyses by status and estimand}
\label{sn:status}

Supplementary Table~\ref{tab:status} distinguishes the estimand, exchangeability unit and analysis status of each comparison.

\begin{table}[h]
\centering\scriptsize
\caption{\textbf{What each analysis estimates, and its status.}}
\label{tab:status}
\begin{tabularx}{\textwidth}{@{}lXll@{}}
\toprule
Analysis & Estimand & Unit of resampling & Status \\
\midrule
H1 & Agreement of the naive population geometry with item wording minus its agreement with self-reported geometry & Participants & Registered (OSF) \\
H2 & Partial correlation of the whitened population geometry with self-reported geometry, beyond wording & Items (permutation) & Registered (OSF), primary \\
H3 & Person-paired item correspondence: mean same-item minus different-item correlation & Participants & Registered (OSF) \\
A1, A2 & Separation of high- from low-quality sessions by stance scores, and its advantage over keyed cosine scores & Sessions & Pre-specified (fingerprinted plan) \\
Psychometric RDM comparisons (Fig.~1) & Second-order agreement of population geometries across sources; split-half reliabilities & Participants (reliability only) & Descriptive \\
General factors, COMPASS PHQ-8 score & Person-paired agreement of language and self-report, alone and beyond length and topic baselines & Participants & Exploratory \\
Person shuffle & Dependence of each statistic on correct person pairing & Shuffles of participants & Exploratory \\
Empirical within-speaker reference & Sensitivity of the H2 partial correlation to the reference covariance; a test of person structure only if within-speaker variation is nuisance shaped like between-speaker nuisance & Items; participants & Exploratory (sensitivity) \\
Keyed alliance scores in psychotherapy & Relation of keyed scores to topic activation & Sessions & Exploratory \\
Probe-bank geometry & Dimensionality and nearest-neighbour structure of item texts & None (descriptive) & Exploratory \\
\bottomrule
\end{tabularx}
\end{table}

\subsection*{Supplementary Note 4: probe banks}
\label{sn:banks}

Supplementary Table~\ref{tab:banks} lists the \nMbBanks{} probe banks. Item texts were transcribed from published forms, manuals or official websites and checked against a second source where available. For BDI-II, PANSS, SANS and SAPS, only available item titles or names were used, without copyrighted response options or definitions. Instruction and stem sentences were excluded. The three versions of the Columbia Suicide Severity Rating Scale were pooled after removing duplicate texts.

\begin{table}[p]
\centering\scriptsize
\caption{\textbf{Probe banks.} Texts were taken from the published forms or manuals; where only item titles or names are openly available, those were used. Copyrighted texts are not redistributed.}
\label{tab:banks}
\begin{tabular}{@{}llrll@{}}
\toprule
Bank & Domain & Items & Text used & Source \\
\midrule
CAPE-42 & symptoms & 42 & items, verbatim & \cite{stefanis2002cape} \\
SAPS & symptoms & 34 & item names & \cite{andreasen1989sans} \\
C-SSRS & symptoms & 31 & items, verbatim & \cite{posner2011cssrs} \\
PANSS & symptoms & 30 & item names & \cite{kay1987panss} \\
SANS & symptoms & 25 & item names & \cite{andreasen1982sans,andreasen1989sans} \\
BDI-II & symptoms & 21 & item titles & \cite{beck1996bdi} \\
PQ-B & symptoms & 21 & items, verbatim & \cite{loewy2011pqb} \\
PCL-5 & symptoms & 20 & items, verbatim & \cite{blevins2015pcl5} \\
BPRS & symptoms & 18 & item names and definitions & \cite{overall1962bprs,guy1976ecdeu} \\
PCL-C & symptoms & 17 & items, verbatim & \cite{blanchard1996pcl} \\
PHQ-15 & symptoms & 15 & items, verbatim & \cite{kroenke2002phq15} \\
HAM-A & symptoms & 14 & item names and definitions & \cite{hamilton1959hama,guy1976ecdeu} \\
AUDIT & symptoms & 10 & items, verbatim & \cite{saunders1993audit} \\
K10 & symptoms & 10 & items, verbatim & \cite{kessler2002k10,kessler2003k6} \\
PHQ-9 & symptoms & 9 & items, verbatim & \cite{kroenke2001phq9} \\
PHQ-8 & symptoms & 8 & items, verbatim & \cite{kroenke2009phq8} \\
GAD-7 & symptoms & 7 & items, verbatim & \cite{spitzer2006gad7} \\
WAI-C & alliance & 36 & items, verbatim & \cite{horvath1989wai} \\
WAI-O & alliance & 36 & items, verbatim & \cite{tracey1989wai} \\
WAI-T & alliance & 36 & items, verbatim & \cite{horvath1989wai} \\
VR-CoDES & interaction & 34 & code definitions & \cite{zimmermann2011vrcodes,delpiccolo2011vrcodes} \\
MITI & interaction & 22 & code definitions & \cite{moyers2016miti} \\
IPIP50 & personality & 50 & items, verbatim & \cite{goldberg1992markers,goldberg2006ipip} \\
TIPI & personality & 10 & items, verbatim & \cite{gosling2003tipi} \\
UCLA-LS3 & process & 20 & items, verbatim & \cite{russell1996ucla} \\
SCS-SF & process & 12 & items, verbatim & \cite{raes2011scssf} \\
EQ-D & process & 11 & items, verbatim & \cite{fresco2007eq} \\
ERQ & process & 10 & items, verbatim & \cite{gross2003erq} \\
PSS-10 & process & 10 & items, verbatim & \cite{cohen1983pss} \\
RRS10 & process & 10 & items, verbatim & \cite{treynor2003rrs} \\
RSES & process & 10 & items, verbatim & \cite{schmitt2005rses} \\
DJG6 & process & 6 & items, verbatim & \cite{dejonggierveld2006djg} \\
WHO-5 & process & 5 & items, verbatim & \cite{topp2015who5} \\
UCLA3 & process & 3 & items, verbatim & \cite{hughes2004ucla3} \\
RDoC & framework & 49 & construct definitions & \cite{insel2010rdoc,cuthbert2013rdoc} \\
\bottomrule
\end{tabular}
\end{table}

\begin{table}[p]
\centering\scriptsize
\caption{\textbf{Entailment hypotheses.} Each symptom item was written as one or more atomic first-person statements; an item's score is the maximum over its statements of $P(\text{entailment})-P(\text{contradiction})$ given the answer window.}
\label{tab:atoms}
\begin{tabularx}{\textwidth}{@{}lX@{}}
\toprule
Item & Hypotheses \\
\midrule
PHQ-8 1 NoInterest & I have little interest in doing things. \newline I have little pleasure in doing things. \\
PHQ-8 2 Depressed & I feel down. \newline I feel depressed. \newline I feel hopeless. \\
PHQ-8 3 Sleep & I have trouble falling asleep. \newline I have trouble staying asleep. \newline I sleep too much. \\
PHQ-8 4 Tired & I feel tired. \newline I have little energy. \\
PHQ-8 5 Appetite & I have a poor appetite. \newline I overeat. \\
PHQ-8 6 Failure & I feel bad about myself. \newline I feel like a failure. \newline I have let myself or my family down. \\
PHQ-8 7 Concentration & I have trouble concentrating on things. \\
PHQ-8 8 Psychomotor & I move or speak so slowly that other people notice. \newline I am fidgety and restless. \\
PCL-C 1 Memories & I have repeated, disturbing memories of a stressful experience. \\
PCL-C 2 Dreams & I have repeated, disturbing dreams of a stressful experience. \\
PCL-C 3 Reliving & I suddenly feel as if a stressful experience were happening again. \\
PCL-C 4 Upset & I feel very upset when something reminds me of a stressful experience. \\
PCL-C 5 Physical & I have physical reactions, like a pounding heart or sweating, when something reminds me of a stressful experience. \\
PCL-C 6 ThoughtAvoidance & I avoid thinking or talking about a stressful experience. \newline I avoid having feelings about a stressful experience. \\
PCL-C 7 ActivityAvoidance & I avoid activities or situations because they remind me of a stressful experience. \\
PCL-C 8 TroubleRemembering & I have trouble remembering important parts of a stressful experience. \\
PCL-C 9 NoInterest & I have lost interest in things that I used to enjoy. \\
PCL-C 10 FeelDistant & I feel distant from other people. \newline I feel cut off from other people. \\
PCL-C 11 FeelNumb & I feel emotionally numb. \newline I am unable to have loving feelings for those close to me. \\
PCL-C 12 Future Cut Short & I feel as if my future will somehow be cut short. \\
PCL-C 13 Sleep & I have trouble falling asleep. \newline I have trouble staying asleep. \\
PCL-C 14 Irritability & I feel irritable. \newline I have angry outbursts. \\
PCL-C 15 Concentration & I have difficulty concentrating. \\
PCL-C 16 HyperAlert & I am super alert or watchful. \newline I am on guard. \\
PCL-C 17 Jumpy & I feel jumpy. \newline I am easily startled. \\
\bottomrule
\end{tabularx}
\end{table}

\clearpage

\clearpage
\bibliographystyle{naturemag}
\bibliography{references}
\end{document}